\documentclass[superscriptaddress,twocolumn,showpacs,preprintnumbers,amsmath,amssymb,prb]{revtex4}

\usepackage{graphicx}% Include figure files
\usepackage{dcolumn}% Align table columns on decimal point
\usepackage{bm}% bold math

\usepackage{natbib}
\usepackage{hyperref} 

\usepackage{color} 
\definecolor{rltred}{rgb}{0.75,0,0}
\definecolor{rltgreen}{rgb}{0,0.6,0}
\definecolor{rltblue}{rgb}{0.3,0.3,1}

\hypersetup{colorlinks,%
    hypertexnames=true,%
    linkcolor=rltgreen,%
    citecolor=rltblue,%
    linktocpage=true}

\begin{document}

\title{Transport through open quantum dots: making semiclassics quantitative}

\author{Iva B\v rezinov\'a}
\email{iva.brezinova@tuwien.ac.at}
\affiliation{Institute for Theoretical Physics, Vienna University of Technology,
Wiedner Hauptstra\ss e 8-10/136, 1040 Vienna, Austria, EU}  

\author{Ludger Wirtz}
\affiliation{Institute for Electronics, Microelectronics, and Nanotechnology
(IEMN), Dept.~ISEN, CNRS-UMR 8520, B.P. 60069, 59652 Villeneuve d'Ascq Cedex, France, EU}

\author{Stefan Rotter}
\affiliation{Institute for Theoretical Physics, Vienna University of Technology,
Wiedner Hauptstra\ss e 8-10/136, 1040 Vienna, Austria, EU}

\author{Christoph Stampfer}
\affiliation{JARA-FIT and II.~Institute of Physics, RWTH Aachen, 52074 Aachen, Germany, EU}

\author{Joachim Burgd\"orfer}
\affiliation{Institute for Theoretical Physics, Vienna University of Technology,
Wiedner Hauptstra\ss e 8-10/136, 1040 Vienna, Austria, EU}  

\date{\today}

\begin{abstract}
We investigate electron transport through clean open quantum dots (``quantum billiards'').
We present a semiclassical theory that allows to accurately reproduce quantum transport calculations. Quantitative agreement is reached for individual energy and magnetic field dependent elements of the scattering matrix. Two key ingredients are essential: (i) inclusion of pseudo-paths which have the topology of linked classical paths resulting from diffraction in addition to classical paths and (ii) a high-level approximation to diffractive scattering. Within this framework of the pseudo-path semiclassical approximation (PSCA), typical shortcomings of semiclassical theories such as violation of the anti-correlation between reflection and transmission and the overestimation of conductance fluctuations are overcome. Beyond its predictive capabilities the PSCA provides deeper insights into the quantum-to-classical crossover.
\end{abstract}


\maketitle
%---------------------------------------------------------------------------
\section{Introduction}
\begin{figure}
	\centering
		\includegraphics[width=5cm]{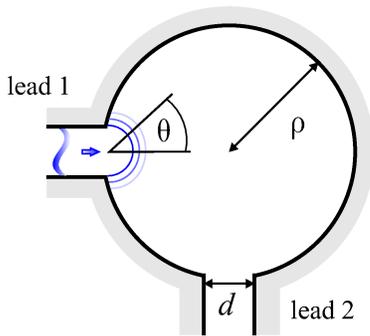}
	\caption{(Color online) Geometry of the circular billiard of radius $\rho$ with perpendicular leads of equal width $d$. In accordance with previous work\cite{IshBur95,RotTanWir00,RotWeiRohBur03} on the circular and the stadium billiard of equal area $a=4+\pi$, we choose $d=0.25$ and $\rho=\sqrt{1+4/\pi}$.}
	\label{fig:circle}
\end{figure}

%---------------------------------------------------------------------------
The ability to controllably fabricate, manipulate and 
examine structures on the sub-micrometer scale has let to the observation of quantum
phenomena in electron transport such as, e.g., universal conductance fluctuations (UCF) in chaotic billiards and weak localization (WL), which
dominate transport at the nanoscale.\cite{Dat95,AkkMon06}
By reducing the characteristic system size below the electronic inelastic mean free path, 
transport enters the so-called ballistic regime.\cite{bee91}
Ballistic electron transport is a prime candidate for semiclassical descriptions~\cite{gutz91,ber72,BraBha03,BluSmi90,BarJalSto93,blom02,SchAlfDel96,BloZoz01,WirTanBur97,WirTanBur99,WirStaRotBur03,StaRotBurWir05,BreStaWirRotBur08,RahBro05,RahBro06,JacWhi06,RicSie02,BraHeuMulHaa06,HeuMulBraHaa06,BroRah06,arg96b,PicJal99,bog00} where the classical 
trajectories carry an amplitude which reflects the stability
of the classical orbits and a phase which contains the classical action and 
accounts for quantum interference.\cite{fey65} On a more fundamental level, the
semiclassical framework provides a conceptually powerful bridge between
classical and quantum mechanics allowing an intuitive approach to quantum mechanics and quantum chaos in general, and to
transport through open quantum dots or so-called quantum billiards in particular.\cite{ber72,gutz91,BraBha03}\\
Several semiclassical approximations (SCAs) based on the approximation of the
constant-energy Green's function for propagation in a billiard 
have been proposed and compared with numerical quantum transport calculations or experiment.\cite{BarJalSto93,blom02,SchAlfDel96,BloZoz01,WirTanBur97,WirTanBur99,WirStaRotBur03,StaRotBurWir05,BreStaWirRotBur08,RahBro05,RahBro06,JacWhi06,ChaBarPfeWes94,MarRimWesHopGos92} While many qualitative features could be well reproduced, quantitative agreement on a system-specific level has remained a challenge.\\
One underlying difficulty is the multi-scale nature of the quantum-to-classical transition for transport through open quantum dots. For the semiclassical approximation to hold, the de Broglie wavelength $\lambda$ should be vanishingly small compared to all characteristic dimensions of the device. Such asymptotic theories have been successfully employed to reproduce, upon ensemble averaging, random matrix theory (RMT) results for chaotic cavities (see e.g.~Ref.~\onlinecite{BluSmi90,RicSie02,BraHeuMulHaa06,HeuMulBraHaa06,BroRah06}). A quantitative comparison on a system-specific level with full quantum calculations or experiments is, however, only possible in the non-asymptotic regime where $\lambda$ is small compared to the linear dimension $D$ of the dot, $\lambda \ll D$, but still comparable to the width of the lead (or quantum wire) $d$, $\lambda \lesssim d$. Moreover, for billiards with sharp edges the proper asymptotic limit is, rigorously, out of reach. The present theory addresses this non-asymptotic semiclassical regime, often also referred to as the ``near'' semiclassical regime. We show that the proper inclusion of diffractive contributions allows to quantitatively reproduce quantum calculations. The diffractive coupling between classical paths gives rise to pseudo-paths that are missing in the standard SCA and are the key to remedy many of the deficiencies of semiclassical approximations.\\
We show in the present communication that this pseudo-path semiclassical approximation (PSCA) can reach quantitative agreement with full quantum simulations provided a high-order diffraction theory for the coupling between classical paths is used. For the scattering at the leads we develop an approximation involving elements of both the uniform theory of diffraction (UTD)\cite{KouPat74,SiePavSch97} and the geometric theory of diffraction (GTD)\cite{Kel62} referred to in the following as the GTD-UTD approximation. With these ingredients good agreement with quantum simulations is found.\\
One key conceptual insight is the unambiguous identification of the paths that contribute to quantum transport. We apply the present theory to a circle-shaped regular quantum dot for which the enumeration of paths, more precisely of path bundles, is easily possible. Unlike for chaotic dots, for which the exponential proliferation of contributing paths as a function of the pathlength makes their unique identification difficult, for regular systems their enumeration and identification is straightforward up to large pathlengths.\\
The circular billiard is depicted in Fig.~\ref{fig:circle}. The leads are attached at right angle and  have equal width $d$. In order to probe the local topology of the cavity we require sufficiently long dwell times such that $d/\rho \ll 1$. The wavelength of the electron $\lambda$ fulfills $\lambda \ll \rho$ for the semiclassical limit to hold inside the cavity. However, as in experimental or numerical studies of quantum billiards $\lambda \lesssim d$. Our semiclassical theory can thus be quantitatively compared with quantum mechanical numerical calculations for the circular billiard.\\
This paper is organized as following: In Sec.~\ref{transport} we review both the standard SCA as well as the PSCA. These approximations differ by the different path sets entering the corresponding Green's function. In Sec.~\ref{sec:diff} we introduce the GTD-UTD diffraction approximation which is a key to the quantitative agreement between the PSCA and quantum mechanics for transport properties. The important role of pseudo-paths and a proper diffraction theory is demonstrated on the level of quantum mechanical length-area spectra \cite{BreStaWirRotBur08} in Sec.~\ref{sec:paths}. Finally, we compare in Sec.~\ref{sec:res} the semiclassical predictions for a variety of quantum transport properties that play a key role in the understanding of the quantum-to-classical crossover, in particular conductance fluctuations (CF), weak localization (WL) and quantum (non-thermal) shot noise, with quantum calculations.
%
%---------------------------------------------------------------------------
\section{Quantum transport through billiards} \label{transport}
%--------------------------------------------------------------------------
We consider ballistic transport through quantum billiards for which transport pro\-perties are determined by the wave number $(k)$ and magnetic field $(B)$ dependent quantum mechanical Hamiltonian $S$-matrix. Dissipative or dephasing processes are neglected. (We will return to the effect of decoherence below.) Moreover, we refer to a ``clean'' billiard when elastic scattering at a disorder potential in the interior of the structure is absent. In this limit, the $S$-matrix represents elastic scattering at the boundaries of the billiard only.\\
The $S$ matrix elements $S_{n,m}^{(j,i)}(k,B)$ describe the scattering from transverse mode $m$
in lead $i$ to mode $n$ in lead $j$. We denote the transmission 
amplitudes from lead $1$ to lead $2$ as $t_{nm}(k,B)=S_{nm}^{(2,1)}(k,B)$
and the reflection amplitudes back into lead 1 as 
$r_{nm}(k,B)=S_{nm}^{(1,1)}(k,B)$. According to the Landauer formula the conductance $g$ of a quantum billiard is directly proportional to the total transmission $T(k,B)$,
\begin{eqnarray}
g(k,B)=\frac{2e^2}{h}T(k,B)=\frac{2e^2}{h}\sum_{n=1}^{M}\sum_{m=1}^{M}|t_{nm}(k,B)|^2,
\label{eq:landcond}
\end{eqnarray}
where $M$ is the number of open modes in the leads having  equal 
width $d$. The $S$ matrix elements can be determined by a projection of the lead modes $\varphi_m(y_i)$ 
($y_i$ is the transverse coordinate in the lead) onto the Green's function $G(y_j,y_i,k,B)$ for 
propagation from $y_i$ to $y_j$ at the cavity lead junction of lead i and lead j, 
respectively. The $S$-matrix elements at $B=0$ are given by the Fisher-Lee equations \cite{FisLee81}
\begin{eqnarray}
& &t_{nm}(k,B=0)=-i\sqrt{k_{x_2,n}k_{x_1,m}}\times \nonumber\\ 
& &\int dy_2\int dy_1 \,\varphi^*_n(y_2)
\,G(y_2,y_1,k,B=0)\,\varphi_m(y_1)\,,
\label{eq:tnm}
\end{eqnarray}
where $k_{x_1,m}$ ($k_{x_2,n}$) is the longitudinal wave number in lead $1$ (lead $2$). The prefactor $\sqrt{k_{x_2,n}k_{x_1,m}}$ is due to flux normalization. We use atomic units 
($\hbar=|e|=m_{\rm eff}=1$). At zero magnetic field the mode wavefunctions take the form
\begin{eqnarray}
\varphi_m(y)=\sqrt{\frac{2}{d}}\sin \frac{m\pi y}{d} \, .
\label{eq:modes}
\end{eqnarray}
For non-zero magnetic field $B\neq 0$, Eqs.~(\ref{eq:tnm}, \ref{eq:modes}) have to be modified (see, e.g., Ref.~\onlinecite{RotTanWir00,RotWeiRohBur03} and references therein). We use the modular recursive Green's function method to calculate the exact quantum mechanical $S$-matrix elements for a given $k$ and $B$ (for details see Ref.~\onlinecite{RotTanWir00,RotWeiRohBur03}).
%----------------------------------------------------------------
\subsection{Semiclassical approximations to transport}
%----------------------------------------------------------------
The starting point of semiclassical approximations are the $S$-matrix elements [Eq.~(\ref{eq:tnm})]. In a first step, the quantum mechanical Green's function $G$ is replaced by a corresponding semiclassical expression which represents the Fourier-Laplace transform in stationary phase approximation of the semiclassical limit of the Feynman propagator. It describes propagation along classical paths with fixed energy. Depending on the class of paths included, a hierarchy of different semiclassical approximations to the Green's function results. These are to be distinguished from the different level of additional approximations employed in the evaluation of the integral [Eq.~(\ref{eq:tnm})] which projects the Green's function onto the lead function. The latter gives rise to another set of semiclassical approximations to the $S$-matrix elements.\\
We focus first on the replacement of $G$ by a semiclassical approximation. For the latter we consider two different levels of approximation, the standard semiclassical approximations (SCA) and the pseudo-path semiclassical approximation (PSCA). Both result from the stationary phase approximation (SPA) to the full Feynman propagator reducing the continuous set of paths entering Feynman's path integral to a discrete subset of paths. Assuming that the classical action $S$ is much larger than $\hbar$ and that well localized and separated stationary points with $\delta S_i = 0$ exist, the standard SCA contains exclusively classical paths. However, near sharp edges of the cavity or near the cavity-lead junctions, the de Broglie wavelength is not negligibly small and the SPA will fail. This leads to diffractive corrections which can be taken into account within the framework of the pseudo-path semiclassical approximation (PSCA). One of its salient features is that the basic notion of a propagator consisting of a sum over a discrete set of paths is preserved. Diffraction effects simply appear as additional contributions to the path sum [see Sec.~\ref{transport} (c)].\\
The classical action for an electron moving along a path 
$q$ is given by $S_q=kL_q+Ba_q/c$, where $L_q$ is the length and $a_q$ is the directed 
enclosed area of the path. $a_q$ can have both positive and negative values depending on the rotational direction of the path. In \emph{all} our semiclassical calculations (standard SCA as well as the PSCA) the magnetic field enters only via the Aharonov-Bohm phase $Ba_q/c$. The curvature of the paths as well as the effect of non-zero magnetic field on the diffraction at the lead (introduced in Sec.~\ref{sec:diff}) is neglected since we consider the regime of weak magnetic fields ($\rho\ll ck/B$ with $\rho$ being the radius of the circular cavity).
%-----------------------------------------------
\subsection{Standard semiclassical approximation}
%-----------------------------------------------
%
\begin{figure}[t]
	\centering
		\includegraphics[width=7cm]{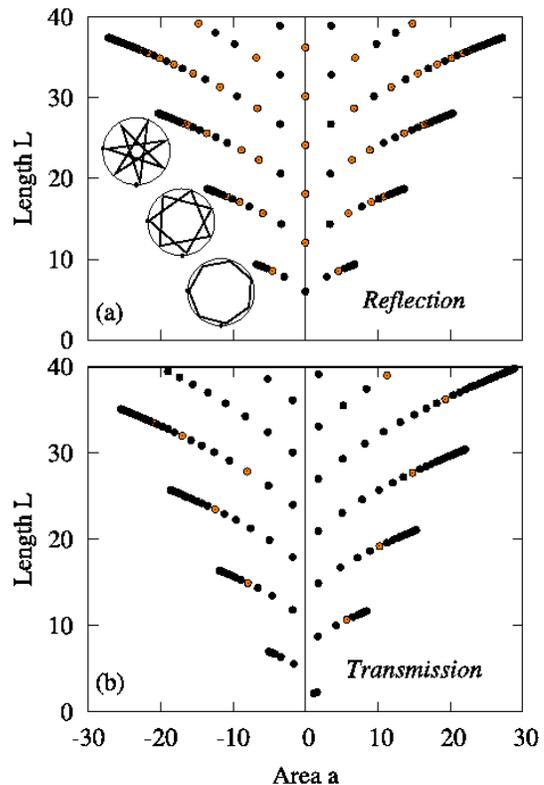}
	\caption{(Color online) Two-dimensional length vs.~enclosed area distribution of classical paths within the open (black dots) and closed (orange dots) circular billiard for (a) reflection and (b) transmission. Each point represents one classical path which connects the centers of each lead. The inset of (a) shows typical paths for the path topology of the three first branches.}
\label{fig:class}
\end{figure} 
The Green's function within the standard SCA entering Eq.~(\ref{eq:tnm}) is given by:
\begin{eqnarray}
G^{\rm SCA}(y_j,y_i,k,B)&=&\frac{2\pi}{{(2\pi i)}^{3/2}}\sum_{q:y_i\rightarrow y_j}\sqrt{|D_q(y_j,y_i,k)|} \nonumber\\&\times&
\exp{\big[iS_q(y_j,y_i,k,B)-i\frac{\pi}{2}\mu_q\big]},
\label{eq:gssca}
\end{eqnarray}
where $D_q(y_j,y_i,k)$ is the deflection factor (a measure for the divergence of nearby 
trajectories) and $\mu_q$ is the Maslov index. 
The deflection factor is defined as
\begin{eqnarray}
|D_q(y_j,y_i,k)|=\frac{1}{|k_{x_j}||k_{x_i}|}\Big|\frac{\partial^2S_q}
{\partial y_j\partial y_i}\Big|
\label{deflec}
\end{eqnarray}
where $x_i$ ($y_i$) is the longitudinal (transverse) component of
the trajectory's starting point ($i$) and end point ($j$), respectively.
The Maslov index increases by two
for every reflection at the hard wall boundary of the 
billiard and by one when passing a focal point along the trajectory. Eq.~(\ref{eq:gssca})
contains a sum over all \emph{classical paths} $q$ connecting the entrance lead $i$ with 
the exit lead $j$ (see Sec.~\ref{sec:claspaths}).\\
Evaluation of Eq.~(\ref{eq:tnm}) with $G^{\rm SCA}$,
\begin{eqnarray}
\label{eq:6}
&&t_{nm}^{\rm SCA} (k, B) = -i \sqrt{k_{x_2,n} k_{x_1,m}}\times\nonumber\\
&&\int\!\!{dy_1} \int\!\!{dy_2}\, \varphi_n^*(y_2) G^{\rm SCA} (y_1, y_2, k, B) \varphi_m (y_1),
\end{eqnarray}
proceeds either numerically\cite{BloZoz01} or analytically by invoking another set of SPAs. It was recognized from very early on that the SPA as applied to Eq.~(\ref{eq:6}) is poorly justified in the non-asymptotic regime when $\lambda\lesssim  d$. Therefore, various diffraction integral approximations have been proposed.\cite{WirTanBur97,SchAlfDel96} The transmission amplitudes then take the form:
\begin{eqnarray}
t_{nm}^{\rm SCA}(k,B)&=&\frac{-1}{\sqrt{2\pi i}}\sqrt{k_{x_2,n}k_{x_1,m}}\sum_{q:y^0_i\rightarrow y^0_j} c_n(\theta_2,k,d)\nonumber\\&\times&
\sqrt{|D_q(k)|}\exp{\big[iS_q(k,B)-i\frac{\pi}{2}\mu_q\big]}\nonumber\\&\times&
c_m(\theta_1,k,d),
\label{eq:tnmssca}
\end{eqnarray}
\noindent expressed in terms of diffraction coefficients $c_m (\theta, k,d)$ describing the diffractive coupling from the entrance lead mode $m$ into the dot and from the dot into the exit lead mode $n$. Deviating from previous calculations, we introduce for $c_{m, n} (\theta, k,d)$ a combination of Keller's geometric
theory of diffraction (GTD) \cite{Kel62} and the ``uniform theory of diffraction'' (UTD).\cite{KouPat74,SiePavSch97} The derivation of $c_{m}(\theta,k,d)$ within the GTD-UTD is given in App.~\ref{appB}. The inclusion of diffraction effects in terms of diffraction coefficients in Eq.~(\ref{eq:tnmssca}) preserves the structure of the semiclassical transmission amplitude in terms of a discrete sum over paths contributing to transport. The diffraction coefficients provide $\theta$- and $k$-dependent weighting factors for each path contributing to the transmission from mode $m$ to mode $n$. Within the framework of the diffractive couplings into and out of the leads, the entrance and exit leads are treated as point scatterers.\cite{StaRotBurWir05} 
Each path bundle which connects the entrance and the exit lead is replaced by an appropriately weighted representative path q connecting the center of the entrance lead with the center of the exit lead. Consequently one can replace the deflection factor as given in Eq.~(\ref{deflec}) by its value in the closed circular billiard $D_q=1/\sqrt{kL_q}$, where $L_q$ is the length of the path.\\
The diffractive lead-dot couplings in Eq.~(\ref{eq:tnmssca}) should be distinguished from diffractive corrections included in the propagation in the interior of the billiard. We refer to Eq.~(\ref{eq:tnmssca}) as the standard SCA while inclusion of diffractive corrections in the billiard corresponds to the PSCA. 
\subsubsection{Paths entering the standard semiclassical Green's function} \label{sec:claspaths}
Classical paths in a regular billiard (such as the circle) feature a highly ordered structure of their length and enclosed area distribution (see Fig.~\ref{fig:class}). The branch structure 
of the ``length-area'' distribution of paths connecting the entrance with the exit point is a specialty of the circular cavity. Along each branch the number of bounces off the wall 
increases by one from one path to its next higher neighbor. The points of convergence of each branch mark those paths that bounce off the wall infinitely many times and thus run exactly 
along the cavity boundary. In the limit $\theta \rightarrow \pi/2$ (where $\theta$ is the entrance angle as given in Fig.~\ref{fig:circle}) each branch contains an infinite 
number of paths. In our numerical calculations (see Sec.~\ref{sec:res}) paths near this cluster point effectively do not contribute as they are cut off by vanishing diffraction coefficients $c_n (\theta \rightarrow \pi/2, k) \rightarrow 0$.\\
The distribution of paths eventually reflected back to the entrance point [Fig.~\ref{fig:class} (a)] is symmetrically distributed relative to the $a=0$ axis due to time-reversal symmetry. Every path has a counterpart of equal length but opposite sign of the enclosed area. The lowest left and right branches consist of polygons with a number of revolutions $n_R=1$ in the cavity. The polygons can be characterized by an angle $\phi=\frac{2\pi}{n_C}$ where 
$n_C$ is the number of corners. Along each branch $n_C$ increases by one from one path to the next. The paths of the next higher branches revolve 
twice ($n_R=2$) around the circle and 
$\phi=\frac{2\pi n_R}{n_C}=\frac{4\pi}{n_C}$. All higher 
branches can be described analogously.\\
The branches of transmitted paths are not symmetric relative to the $a = 0$ axis but show a clear off-set (Fig.~\ref{fig:class} (b)). Path pairs with similar length do not have, in general, the same topology but differ in the number of bounces off the hard wall boundary. As a consequence, these path pairs have different Maslov indices and thus do not interfere constructively. Fig.~\ref{fig:class} also illustrates the difference between the open and closed billiard, i.e., the effect of ``path shadowing'' or suppression of longer paths due to their prior exit from the structure. All paths that would be geometrically reflected off the closed lead in the closed system are missing in the open billiard. The difference between the closed and open billiard is particularly evident in reflection since all paths with four-fold symmetry leave the cavity via lead 2 before being reflected back to lead 1. This is system-specific for the circular billiard with perpendicular leads (see Fig.~\ref{fig:circle}).\\
The distinctly different path distributions for transmission and reflection point already to a clear structural deficiency of the SCA. Many quantum properties of transport are a consequence of the intrinsic coupling between transmission and reflection. The standard SCA does not incorporate this quantum aspect of non-locality. Classically, the path sets of transmission and reflection are disjoint. The distribution of transmitted paths is markedly different from the one of reflected paths. In quantum transport transmitted and reflected paths are intertwined and must share the information on the relative phases. A semiclassical theory that reproduces quantum features must therefore allow for coupling between the path sets associated with transmission and reflection. This is the key property of pseudo-paths discussed in the following.
%
%-------------------------------------------------------------------------------------
\subsection{Pseudo-path semiclassical approximation} \label{se:psca}
%-------------------------------------------------------------------------------------
%
The pseudo-path semiclassical approximation~\cite{WirStaRotBur03,StaRotBurWir05} goes beyond the standard SCA by systematically including diffractive corrections into the propagation in the interior of the billiard. In the present case, diffractive corrections arise from multiple back scattering, i.e., internal reflections at the leads. We point to the conceptual similarity to ``pseudo-orbits'' introduced in Ref.~\onlinecite{SzeGoo93} for the study of the density of states in a closed wedge billiard and to ``Hikami boxes''\cite{Hik81,AkkMon06} introduced for elastic scattering at short-ranged potentials in the interior of a diffusive quantum dot.\\
In line
with multiple scattering theory the pseudo-path semiclassical Green's function can be expressed in terms
of a (semiclassical) Dyson equation,\cite{StaRotBurWir05}
\begin{eqnarray}
G^{\rm PSCA}& =& G^{\rm SCA}+G^{\rm SCA} V G^{\rm PSCA}  \nonumber \\
 & = & G^{\rm SCA} \sum_{i=0}^{\infty}(V G^{\rm SCA})^i.
\label{dysonpsc}
\end{eqnarray}
In the present case, $G^{\rm SCA}$ plays the role of the unperturbed
Green's function and the perturbation ``potential'' $V$ accounts for the
(internal) diffractive scatterings at the lead opening. The unperturbed propagation inside the
cavity $G^{SCA}$ is equivalent to the free propagation in
the ``closed'' system.\\
For a two-terminal system the 
perturbation potential $V$ is given by
\begin{eqnarray}
V  =  {V_1 \,\, 0 \choose 0 \,\, V_2},
\end{eqnarray}
where $V_{1}$ and $V_{2}$ describe the diffractive scattering
off lead 1 and lead 2, respectively. The semiclassical Dyson equation, Eq.~(\ref{dysonpsc}), reads
\begin{eqnarray}
\label{dysmat}
& &{G_{1,1}^{\rm PSCA} \,\, G_{1,2}^{\rm PSCA} \choose G_{2,1}^{\rm PSCA}
 \,\, G_{2,2}^{\rm PSCA}}  =  {G_{1,1}^{\rm SCA} \,\, G_{1,2}^{\rm SCA} \choose G_{2,1}^{\rm SCA} \,\, G_{2,2}^{\rm SCA}} \nonumber \\
& &+{G_{1,1}^{\rm SCA}  \,\, G_{1,2}^{\rm SCA} \choose G_{2,1}^{\rm SCA} \,\, G _{2,2}^{\rm SCA}} 
{V_1 \,\, 0 \choose 0 \,\, V_2}
{G_{1,1}^{\rm PSCA} \,\, G_{1,2}^{\rm PSCA} \choose G_{2,1}^{\rm PSCA} \,\, G_{2,2}^{\rm PSCA}}.
\end{eqnarray}
To first order, the PSCA to the Green's function (denoted by $G^{\rm PSCA (1)}$) connecting lead $i$ with lead $j$ includes terms in $V$ of the form,
\begin{eqnarray}
G^{\rm PSCA (1)} &=& \sum_{l=1,2}G_{j,l}^{\rm SCA} V_l G_{l,i}^{\rm SCA}\nonumber \\
 &=&\sum_{l=1,2} \sum_{q'_{j,l}} \sum_{q_{l,i}} G_{q'_{j,l}}^{\rm SCA} v(\theta_{q'_l},\theta_{q_l},k,d) G_{q_{l,i}}^{\rm SCA}. \nonumber \\
\label{eq:11}
\end{eqnarray}
Eq.~(\ref{eq:11}) may serve as example to illustrate the physics entering Eq.~(\ref{dysmat}). It describes propagation from lead $i$ to lead $j$ via one intermediate visit to lead $l$ where diffractive internal back scattering with amplitude $v(\theta_{q'_l},\theta_{q_l},k,d)$ takes place. The path from lead $i$ to lead $l$, $q_{l,i}$ as well as from lead $l$ to lead $j$, $q'_{j, l}$, are classical paths described by $G^{\rm SCA}$. Diffractive scattering couples the incident path $q_{l,i}$ with angle $\theta_{q_l}$ to the exiting path $q'_{j, l}$ with angle $\theta_{q'_l}$ thereby coupling two disjoint subsets of classical paths and generating a first-order pseudo-path. The determination of the diffraction coefficient $v (\theta_{q'_l},\theta_{q_l},k,d)$ will be discussed in more detail in Sec.~\ref{sec:diff}. From a conceptual point of view, the pseudo-path semiclassical approximation [Eq.~(\ref{dysmat})] is closely related to the diagrammatic perturbation theory.\cite{Dat95} Both leads 1 and 2 (i.e., $V_{1}$ and $V_{2}$)
act much like ``Hikami boxes'',\cite{AkkMon06,Hik81} where electrons cannot be described semiclassically since the characteristic
potential length scale (in our case the sharp edges of the leads) is smaller than the electron wavelength $\lambda$.
Thus, the wave nature of electrons has to be taken into account as diffraction at the point scatterers allowing classically distinct path sets to mix. This is crucial for the quantum corrections to the transmission and reflection amplitudes and repairs some of the deficiencies of the standard SCA (see Sec.~\ref{sec:res}).
%
%-----------------------------------------------------------------------------
\subsubsection{Paths entering the pseudo-path semiclassical Green's function}
%-----------------------------------------------------------------------------
%
\begin{figure}[t]
	\centering
		\includegraphics[width=7cm]{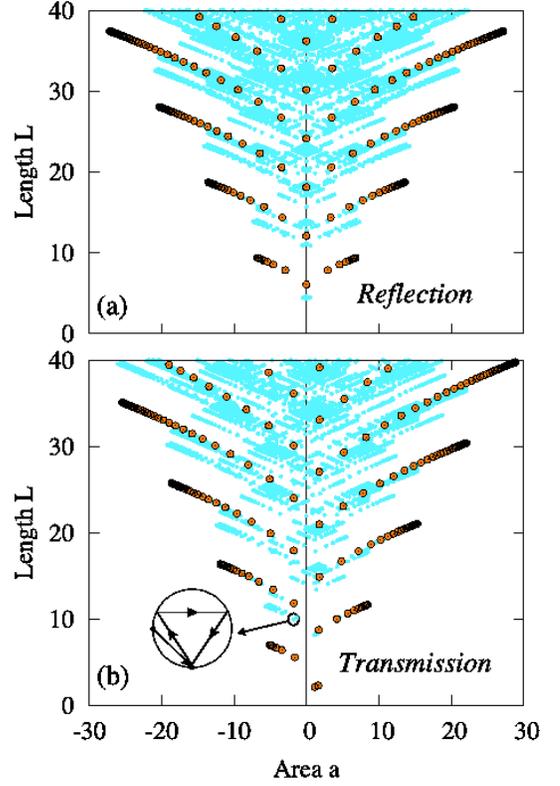}
	\caption{(Color online) Length-area distribution of first-order pseudo-paths (blue dots) for reflection (a) and transmission (b). Zeroth-order pseudo-paths (i.e., the classical paths) are denoted by orange dots (same as Fig.~\ref{fig:class}). The number of first-order pseudo-paths up to length $L = 40$ is larger by more than a factor of $60$ compared to the number of classical paths.}
\label{fig:class1kink}
\end{figure}
Within the PSCA pseudo-paths are formed by joining 
classical paths together via diffraction. With each increasing order of the PSCA the length-area distribution gets more
and more densely filled with paths, or equivalently, the number of paths increases with the order of the PSCA (Fig.~\ref{fig:class1kink}). Pseudo-paths form product sets of classical paths, e.g., joining a given classical path for transmission with a classical path contributing to reflection (or in reverse order) forms the set of transmitted pseudo-paths to first order. Reflected first-order pseudo-paths result from joining two classical transmitted paths or two classical reflected paths. This coupling allows to recover the non-locality of quantum transport. Higher-order pseudo-paths are constructed analogously. With increasing order and increasing (combined) pathlength the total number of pseudo-paths exponentially proliferates. This is in sharp contrast to the power-law growth of purely classical paths for regular systems and explains why the effect of diffractive scattering is more likely visible in regular than in chaotic systems where the exponential proliferation of classical orbits may mask the diffractive contributions.\\
The length-area distribution of first order pseudo-paths contained in Eq.~(\ref{dysmat}) contributing to reflection is (of course) still symmetric [Fig.~\ref{fig:class1kink} (a)]. In addition, new branches appear with classical and pseudo-path partners of approximately equal length. The change in the branch structure is more dramatic in the spectrum of transmitted paths [Fig.~\ref{fig:class1kink} (b)]. The pseudo-paths lead to a ``symmetrization'' of the length-area distribution. The symmetrization results primarily from paths which \emph{change} their \emph{rotational direction} through diffractive scattering [see the pseudo-path in the inset of Fig.~\ref{fig:class1kink} (b)]. Branches of classical paths are now completed by symmetric pseudo-path ``partner'' branches of approximately equal length and different enclosed area.\\
The weight and the phase of interfering paths are strongly influenced by the diffraction coefficient $v(\theta',\theta,k,d)$. As will be demonstrated in Sec.~\ref{sec:res}, the previously employed Fraunhofer theory of diffraction \cite{WirTanBur97,WirTanBur99,WirStaRotBur03,StaRotBurWir05} is not sufficiently accurate as to give quantitatively reliable results for transport properties. The same holds for the Kirchhoff theory of diffraction \cite{SchAlfDel96} which is closely related to the Fraunhofer theory of diffraction and gives similar results for the diffraction coefficients (see Fig.~\ref{fig:diff_qm} in the following section). For a quantitative agreement between the PSCA and quantum mechanics, it is thus necessary to go beyond low-order diffraction approximations and use a more sophisticated theory of diffraction. We present such a theory in the following section.
%
%---------------------------------------------------------------------------
\section{Diffraction at the lead} \label{sec:diff}
%---------------------------------------------------------------------------
%
\begin{figure}[t]
	\centering
		\includegraphics[width=7cm]{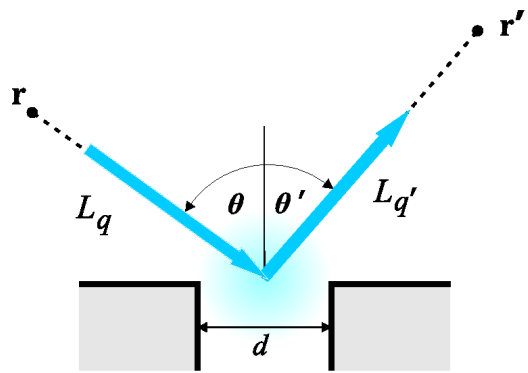}
	\caption{(Color online) Sketch of diffraction at an open lead mouth:
a path $q$ reaches the orifice under an angle $\theta$ and is
backscattered into a path $q'$ that leaves with an angle $\theta'$ (the angles $\theta$ and $\theta'$ as depicted in the figure have opposite signs). The dashed lines denote that $\vec{r}$ and $\vec{r}\,'$ are in the far field region.}
\label{fig:diff_sketch}
\end{figure}
\begin{figure}
	\centering
		\includegraphics[angle=-90,width=8cm]{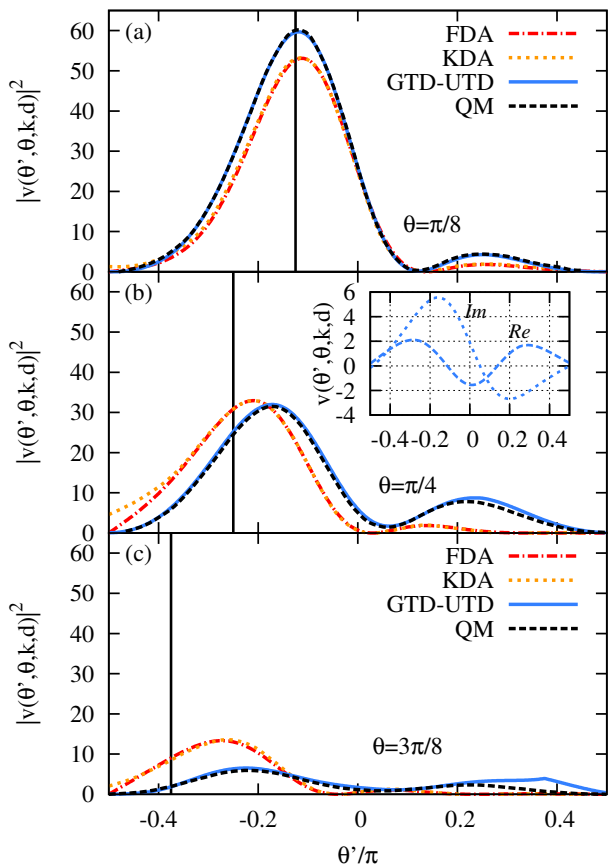}
	\caption{(Color online) Absolute square of the diffraction coefficient 
$|v(\theta',\theta,k,d)|^2$ for diffractive scattering (see Fig.~\ref{fig:diff_sketch}) within the FDA, the KDA, the GTD-UTD and exact quantum mechanical data for angle of incidence (a) $\theta=\pi/8$, (b) $\theta=\pi/4$, and (c) $\theta=3\pi/8$ 
at $k=2.5\pi/d$. The vertical lines in each frame mark the angles of specular reflection.
The quantum mechanical (QM) and KDA coefficients
are taken from Ref.~\onlinecite{SchAlfDel96}. Inset of 
Fig.~(b): the real and imaginary part of $v(\theta',\theta,k,d)$ within the GTD-UTD.}
\label{fig:diff_qm}
\end{figure}
The contribution of a given classical path with pathlength $L_q$ to the
standard semiclassical Green's function, $G^{\rm SCA}$, is
\begin{eqnarray}
G^{\rm SCA}_q =\frac{1}{\sqrt{2\pi kL_q}}e^{ikL_q-i3\pi/4-i\frac{\pi}{2}\mu_q} \, .
\label{eq:gfree}
\end{eqnarray}
Eq.~(\ref{eq:gfree}) is the basic building block entering the Dyson Eq.~(\ref{dysmat}) together with the diffraction coefficient $v (\theta', \theta, k, d)$. In line with the far-field approximation underlying diffraction theory, $v (\theta', \theta, k, d)$ is assumed to be independent of the length of the path approaching or exiting the diffractive scattering region.\\
A successful application of the pseudo-path
semiclassical approximation outlined in the preceding subsection
requires the determination of accurate diffraction coefficients
$v(\theta',\theta,k,d)$. Different approximations have
been used in the past for the inclusion of diffraction effects:
Kirchhof diffraction approximation (KDA),\cite{SchAlfDel96} 
Fraunhofer diffraction approximation (FDA),\cite{WirTanBur97,WirTanBur99,WirStaRotBur03,StaRotBurWir05} 
geometric theory of diffraction (GTD),\cite{Kel62}
and the ``uniform theory of diffraction'' (UTD).\cite{KouPat74,SiePavSch97}
We have developed a new theory for the reflection at open lead mouths by combining the GTD with the UTD (the GTD-UTD) to take into account paths that scatter multiple times between the 
edges of the leads, (see appendix~\ref{appA}). With this theory we obtain diffraction coefficients in excellent agreement to quantum mechanics.\\
Consider, as a test case, the diffractive scattering  (Fig.~\ref{fig:diff_sketch}) at the lead mouth described by the first-order term [Eq.~(\ref{eq:11})],
\begin{eqnarray}
\label{eq:PSCA}
G_{q', q}^{\rm PSCA^{(1)}} = G_{q'}^{\rm SCA} v(\theta', \theta, k,d) G_q^{\rm SCA} \, .
\end{eqnarray}
The incoming path $q$ is incident at angle $\theta$ (measured with respect to the surface normal) and is diffractively scattered into angle $\theta'$ under which path $q'$ leaves the scattering region. We compare (Fig.~\ref{fig:diff_qm}) the present GTD-UTD theory with the FDA, the KDA and exact quantum mechanical (QM) calculations.\cite{SchAlfDel96} Even for a typical $k$ value in the low mode regime ($k=$2.5$\pi/d$), the agreement between the GTD-UTD and the exact QM calculations is very good whereas both the FDA and the KDA display clear deviations from the QM values. While these deviations do not appear dramatic at first glance, they are, in fact, quantitatively very important as the diffraction coefficient enters the Dyson series [Eq.~(\ref{dysmat})] to all orders. Note, however, that the GTD-UTD would fail in the limit $\theta \rightarrow \pi/2$ [as indicated by the kink at $\theta' = \theta=3\pi/8$ in Fig.~\ref{fig:diff_qm} (c)]. This deficiency is of no concern for the present applications as the probability for diffractive scattering tends to zero in this limit. The maximum of the diffractive reflection probability is clearly around the specular value $\theta' \approx -\theta$. However, it is important
to note that the probability distribution possesses a local maximum 
at the backscattering angle $\theta' \approx \theta$. A non-negligible part of the
electron wave is back-scattered into the direction from where it came from. This diffractive back-reflection should not be confused with the well-known Andreev back-reflection in which the back-reflected particle simultaneously undergoes a particle-hole exchange.\cite{And64} Back-reflected paths are responsible for the symmetrization of the distribution of transmitted paths [Fig.~\ref{fig:class1kink} (b)] and are crucial for the understanding of the weak-localization dip in the 
transmission.\cite{BreStaWirRotBur08}\\
While the KDA and the FDA have been successfully
used in the past to explain certain features of conductance
fluctuations,\cite{SchAlfDel96,WirTanBur97,WirTanBur99,WirStaRotBur03,StaRotBurWir05} only the GTD-UTD 
is precise enough to reproduce transport semiclassically on a quantitative level (see Sec.~\ref{sec:res}). In Sec.~\ref{sec:wl} we compare the results for transport properties obtained by implementing the GTD-UTD and the FDA to full quantum mechanical calculations.   
%
%---------------------------------------------------------------------------
\section{Paths in quantum transport}\label{sec:paths}
%---------------------------------------------------------------------------
%
\begin{figure}[t]
	\centering
		\includegraphics[angle=-90,width=8.5cm]{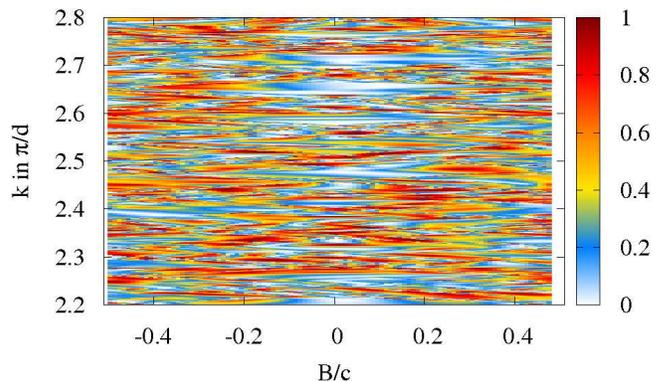}
	\caption{(Color online) Two-dimensional distribution of the absolute square of the quantum $S$-matrix element $|t_{22}(k,B)|^2=T_{22}(k,B)$ as a function of the wavenumber $k$ and the magnetic field $B$.}
\label{fig:t22kB}
\end{figure}
\begin{figure*}[t]
	\centering
		\includegraphics[angle=-90,width=17cm]{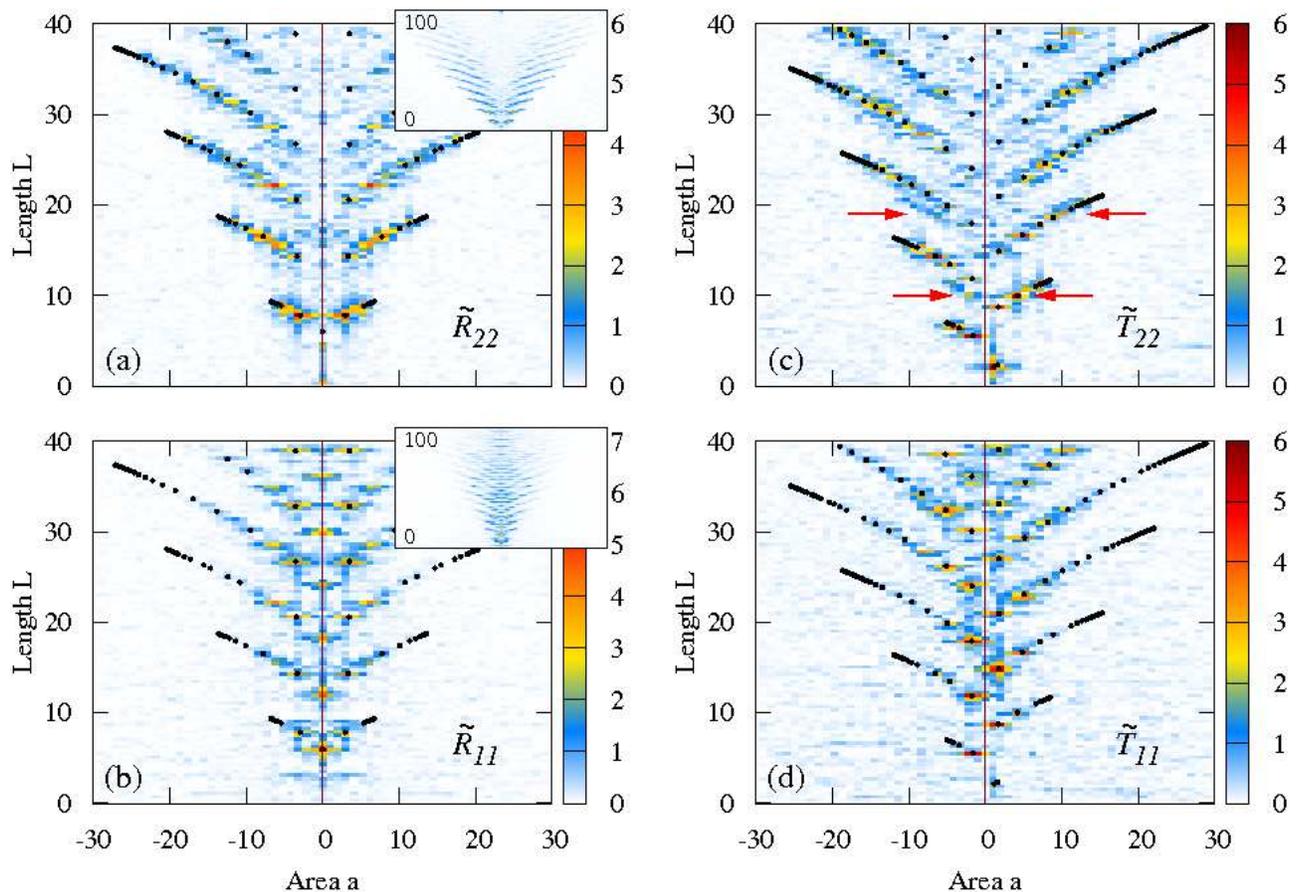}
	\caption{(Color online) The absolute square of the Fourier transform $|\tilde{S}(L,a)|^2$ as a function of length $L$ and enclosed directed area $a$: (a) $\tilde{R}_{22}(L,a)$, (b) $\tilde{R}_{11}(L,a)$, (c) $\tilde{T}_{22}(L,a)$, and (d) $\tilde{T}_{11}(L,a)$. The color shading is determined by $\log{(|\tilde{S}(L,a)|^2+1)}$. The insets show the probabilities along the entire resolved length. The integration of Eq.~(\ref{eq:fouriert}) is performed numerically over $k\in [2.2,3.45]\pi/d$ discretized with $251$ points and $B/c\in[-3,3]$ with $121$ points. The spectra are compared to the distribution of classical paths (black dots) at zero $B$ field (the curvature of the trajectories in the present magnetic field and energy regime is negligible).}
\label{fig:laspec}
\end{figure*}
\begin{figure*}
	\centering		\includegraphics[angle=-90,width=17cm]{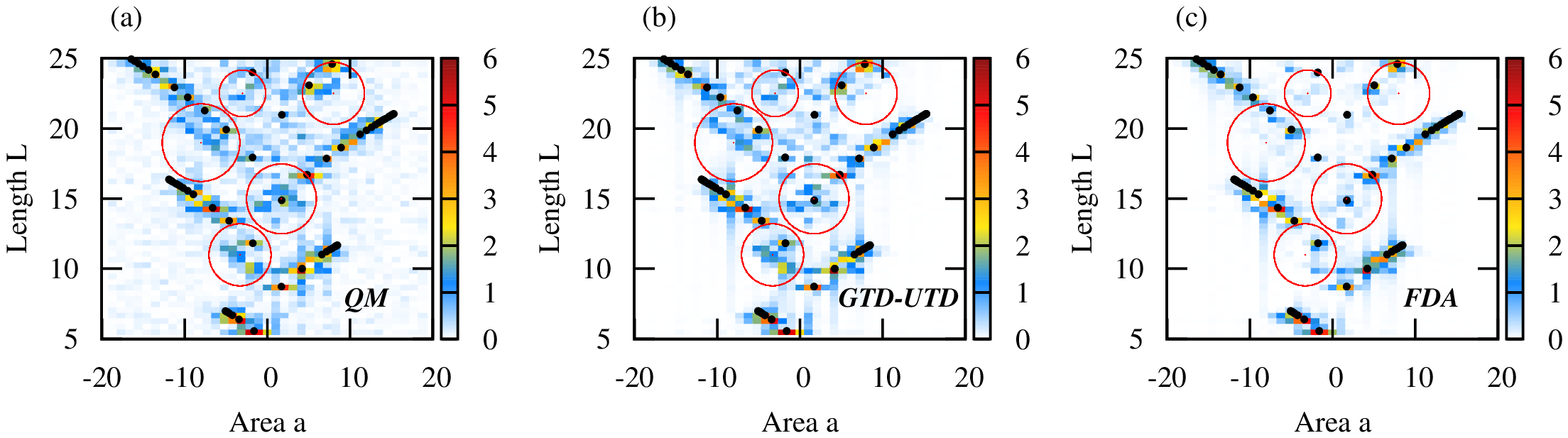}
	\caption{(Color online) $\tilde{T}_{22}(L,a)$ as a function of length $L$ and enclosed directed area $a$: (a) full quantum $S$-matrix, (b) the PSCA using the GTD-UTD, and (c) the PSCA using the FDA for the diffraction coefficients. Several regions with non-classical paths are highlighted by red circles.}
\label{fig:laspec_comp}
\end{figure*}
The information on paths governing quantum transport can be reliably extracted from the two-dimensional Fourier transforms of the quantum mechanical $S$-matrix elements,\cite{BreStaWirRotBur08} $S_{nm}(k,B)$. The $S$-matrix elements display a strongly fluctuating pattern as a function of $k$ and $B$ (see Fig.~\ref{fig:t22kB} for $T_{22}(k,B)$). Since the canonically conjugate variables to the 
wavenumber $k$ and the magnetic field $B$ are the length $L$ and the directed area $a$, respectively, the Fourier transform
\begin{eqnarray}
\tilde{S}^{j,i}_{nm}(L,a)=\int \! \! dk\int \! \! dB\ e^{-i(kL+\frac{B}{c}a)}S^{j,i}_{nm}(k,B)
\label{eq:fouriert}
\end{eqnarray}
allows the unambiguous identification of quantum paths contributing to quantum transport via their length and enclosed area [Fig.~\ref{fig:laspec} (a)-(d)]. No apriori-assumption as to the existence of classical paths $q(L,a)$ with pathlength $L$ and area $a$ enters Eq.~(\ref{eq:fouriert}). The two-dimensional pathlength-area spectrum allows to identify both classical as well as non-classical contributions to the full quantum spectra. Fig.~\ref{fig:laspec} displays examples of pathlength-area spectra $|\tilde{S}^{i j}(L, a)|^2$. [Note that Fig.~\ref{fig:laspec} (c) is the absolute square of the Fourier transform of the transmission amplitude whose absolute square is plotted in Fig.~\ref{fig:t22kB}]. Obviously the strong fluctuations of conductance in quantum transport (Fig.~\ref{fig:t22kB}) are the result of the interference of clearly identifiable (quantum) paths (Fig.~\ref{fig:laspec}).\\
The quantum mechanical $S$-matrix elements $S_{nm}(k,B)$ are determined with the help of the modular recursive Green's function method\cite{RotTanWir00,RotWeiRohBur03} and then numerically 
Fourier transformed. Finite discretized intervals must be used when performing the Fourier transform numerically. The integration intervals are denoted by 
$\Delta k=k_{max}-k_{min}$ ($\Delta B=B_{max}-B_{min}$) and the numerical grid spacings by $\delta k$ ($\delta B$). Accordingly, the resolvable 
length is $\Delta L$ = $2\pi/\delta k$ and the resolvable area interval is 
 $\Delta a = 2\pi/(\delta B/c)$. The magnitude of the $S$-matrix elements decreases with increasing 
length, which is an obvious consequence of open systems: the probability to stay within the 
cavity decreases with increasing length. The parameter $\delta k$ must be chosen 
sufficiently small such that the maximum resolvable length ($\Delta L$) lies already in the region of 
strongly reduced amplitudes. Otherwise, the Fourier spectrum is visibly back-folded onto the fundamental interval. The magnetic field interval is further restricted by the requirement that the curvature of the 
paths is negligible, i.e., the cyclotron radius $ck/B$ should be much larger than
the circle radius $\rho$. We chose the interval $\Delta k$ and $\Delta B$ as well as the number 
of interval points such that a maximum length of $\Delta L = 100$ is resolved 
[in Fig.~\ref{fig:laspec} (a)-(d) only the contributions with $L\le40$ are shown, the insets contain the 
entire spectrum]. Except for $R_{11}$, the absolute square of the $S$-matrix elements is
considerably damped at a length of $\Delta L=100$. Thus, the graphs
represent essentially the entire length-area spectrum. Only for $R_{11}$,
contributions with $L>100$ are non-negligible which leads
to a back-folding near $L=0$ [the lowest branch structure near $L=2$ in Fig.~\ref{fig:laspec} (b)].\\
The quantum length-area spectra provide detailed information on the paths contributing to transport through a specific system. They represent the paths entering the \emph{full} Feynman path integral. The following general trends can be observed: Long paths are more prevalent in $S$-matrix elements connecting low mode numbers (in the present case they contribute more strongly to $S$-matrix elements $\tilde{S}_{11}$ than to $\tilde{S}_{22}$). Lower modes favor smaller entrance and exit angles that are associated with longer paths with a larger number of bounces off the cavity boundary. The most important observation is the remarkably close correspondence of the quantum mechanical length-area spectrum to its classical counterpart. Important contributions are located near classical paths. Moreover the branch structure of classical paths is reproduced [Fig.~\ref{fig:laspec} (a)-(d)]. On the other hand, there are distinct structures 
which do not correspond to classical paths and which can be identified using the distributions in Fig.~\ref{fig:class1kink} (a) and (b). The quantum mechanical 
length-area spectra confirm the existence and substantial role of non-classical paths, the pseudo-paths: these are those paths that are one or several times diffractively reflected off the lead before exiting the cavity. Two examples are given in the following: For $\tilde{T}_{22}$
we identify an ensemble of diffractive paths which, among others, contribute to a symmetrization of the
spectrum [non-classical branches in Fig.~\ref{fig:laspec} (c), two of them together with the classical partner branches are marked by arrows]. $R_{11}$ reveals the importance of paths that are geometrically reflected off the open lead (e.g., the periodic contributions near $a=0$ along the length axis belong to horizontal paths bouncing increasingly many times back and forth).\\
The importance of a given class of paths to quantum transport can be delineated by inverting this decomposition process. Deleting a selected class of paths (classical or non-classical) from $\tilde{S}_{ij} (L, a)$ and performing the inverse Fourier transform gives rise to truncated $S$-matrix elements $\bar{S}_{ij} (k, B)$ from which certain path contributions have been removed in a controlled manner. This is the key to detailed quantitative tests of semiclassical theories. Since summation of the (P)SCA over arbitrarily long paths is prohibitively complicated we can compare truncated quantum and semiclassical $S$-matrix elements where both quantum and classical paths only up to a maximum pathlength $L \leq L_{max}$ are included. The length-area spectra also allow sensitive tests for the proper diffractive weight of a given class of paths in a semiclassical theory. To this end, we first calculate the $S$-matrix elements within the PSCA and then perform the Fourier transform [Eq.~(\ref{eq:fouriert})] in analogy to quantum calculations. To analyze the role of a proper diffraction 
coefficient we use either the GTD-UTD (which gives good agreement with quantum mechanics, see Fig.~\ref{fig:diff_qm}) or the FDA (with poor agreement with quantum mechanics, see likewise Fig.~\ref{fig:diff_qm}). The fact that back-reflection into the cavity is poorly described within the FDA is
mirrored in the semiclassical length-area spectrum where important non-classical (diffractive) contributions have a far
too low weight [Fig.~\ref{fig:laspec_comp} (c)]. In particular, the diffractive change of the rotational direction is insufficiently described (see Fig.~\ref{fig:diff_qm} for $\theta'>0$). A clear indication for the essential role of the corresponding paths is the improvement within the GTD-UTD. The length-area spectrum within the GTD-UTD remarkably reproduces even fine details of the quantum mechanical spectrum [compare Fig.~\ref{fig:laspec_comp} (a) and (b)]. The non-classical path sets in Fig.~\ref{fig:laspec_comp} can be identified using the path distributions within the PSCA (Fig.~\ref{fig:class1kink}). The length-area spectra do not leave any ambiguity as to which paths contribute by which phase and weight. By taking into account pseudo-paths and weighting them with the appropriate diffraction coefficients the quantum mechanical path spectrum is reproduced on a quantitative level. In the following section we demonstrate how the accurate representation of the length-area spectra within PSCA directly translates into quantitative reproduction of transport properties.
%\FloatBarrier
%---------------------------------------------------------------------------
\section{Application to transport through regular billiards} \label{sec:res}
%---------------------------------------------------------------------------
Phase coherent ballistic transport governed by quantum interference  influences the conductance in several important ways: the conductance strongly fluctuates as a function of the Fermi energy, the magnetic field or the cavity geometry (conductance fluctuations, CF).  The conductance is, on average, suppressed compared to the classical prediction and increases with external magnetic field (weak localization, WL). The noise carries the signatures of quantum mechanical uncertainty (shot noise).\\
Path interference has been the key to the understanding of phase coherent ballistic transport.\cite{BluSmi90,BarJalSto93,blom02,SchAlfDel96,BloZoz01,WirTanBur97,WirTanBur99,WirStaRotBur03,StaRotBurWir05,BreStaWirRotBur08,RahBro05,RahBro06,JacWhi06,RicSie02,BraHeuMulHaa06,HeuMulBraHaa06,BroRah06,arg96b,PicJal99,bog00,yan95} For chaotic systems and large mode numbers quantum transport properties have been attributed \cite{RahBro05,RahBro06,JacWhi06,RicSie02,BraHeuMulHaa06,HeuMulBraHaa06,BroRah06} to the interference of classically allowed paths (SCA). We demonstrate in the following for the circular billiard at low mode numbers (accessible to experiments) that CF, WL, and shot noise cannot be explained by the interference of classical paths alone. In this regime, the SCA overestimates the CF, does not reproduce the weak-localization dip and also shows poor quantitative agreement with the quantum mechanical prediction for shot noise. These difficulties can be overcome with the PSCA when a high-level diffraction approximation, the GTD-UTD, is employed. For technical reasons, we perform in both the SCA and the PSCA the summation over paths only up to a maximum cut-off length $L_{\rm max}$. Correspondingly, we truncate the full quantum scattering matrix elements by setting all elements $\tilde{S}_{nm}(L, a) $ with lengths exceeding $L_{\rm max}$ to 0 and carry out the inverse Fourier transform. This allows a quantitative comparison between semiclassical and quantum calculations unaffected by the (inevitable) unitarity deficiency of a truncated semiclassical path sum.\\
%
%---------------------------------------------------------------------------
\subsection{Conductance fluctuations} \label{sec:condfl}
%---------------------------------------------------------------------------
%
\begin{figure*}
	\centering
		\includegraphics[angle=-90,width=17cm]{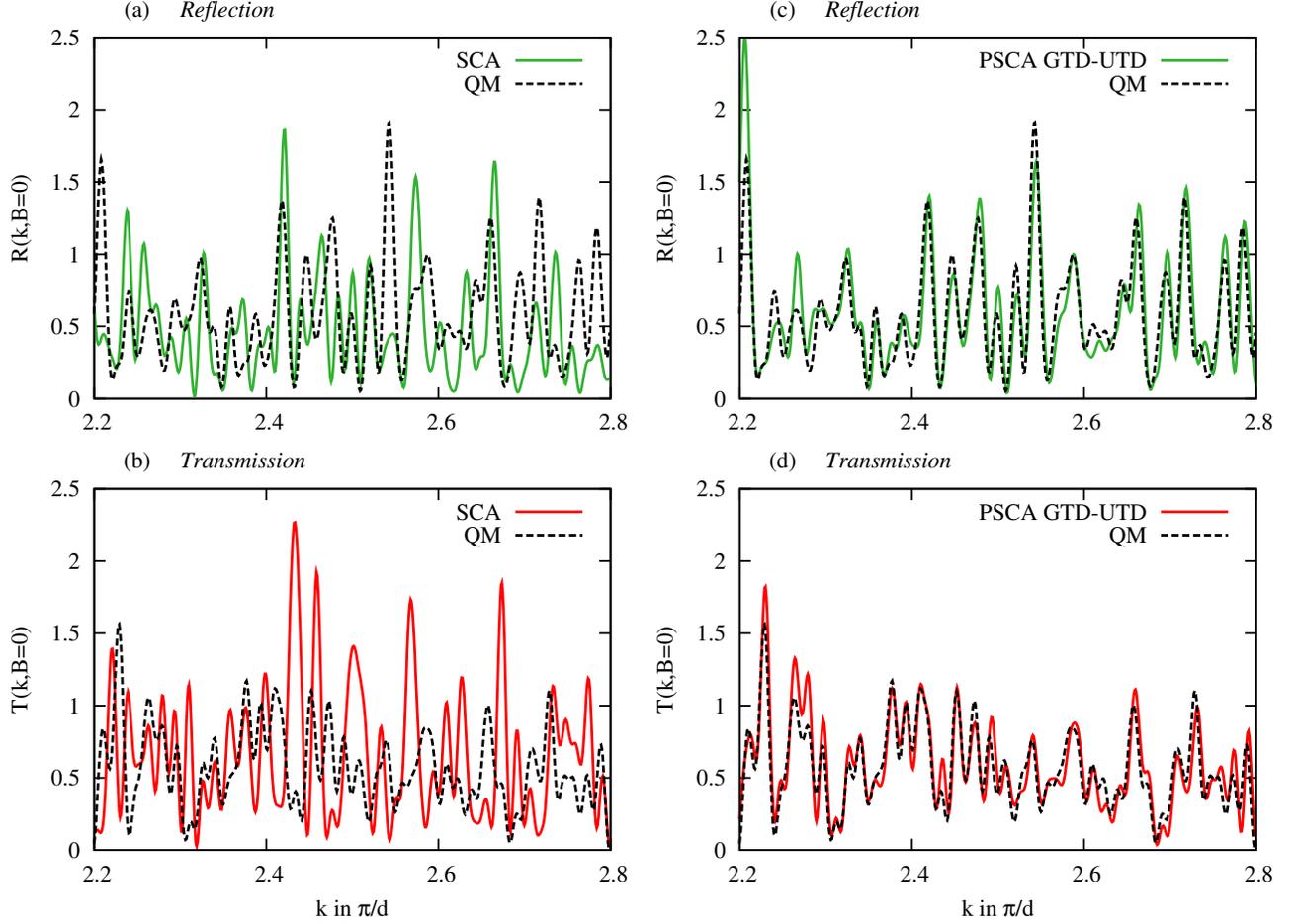}
	\caption{(Color online) (Conductance) Fluctuations in the transmission $T(k,B=0)$ and reflection $R(k,B=0)$ as a function of $k$. The maximal included length is $L_{\rm max}=40$. (a) Reflection and (b) transmission within the SCA (red/green solid line), and quantum mechanics (black dashed line). For the diffraction coefficients $c_m(\theta,k,d)$ entering the SCA we have used the GTD-UTD. (c) Reflection and (d) transmission within the PSCA with the GTD-UTD (red/green solid line), and quantum mechanics (black dashed line). The PSCA includes diffractive scattering up to $\rm{5^{th}}$ order.}
\label{fig:kdep}
\end{figure*}
\begin{figure*}
	\centering
		\includegraphics[angle=-90,width=17cm]{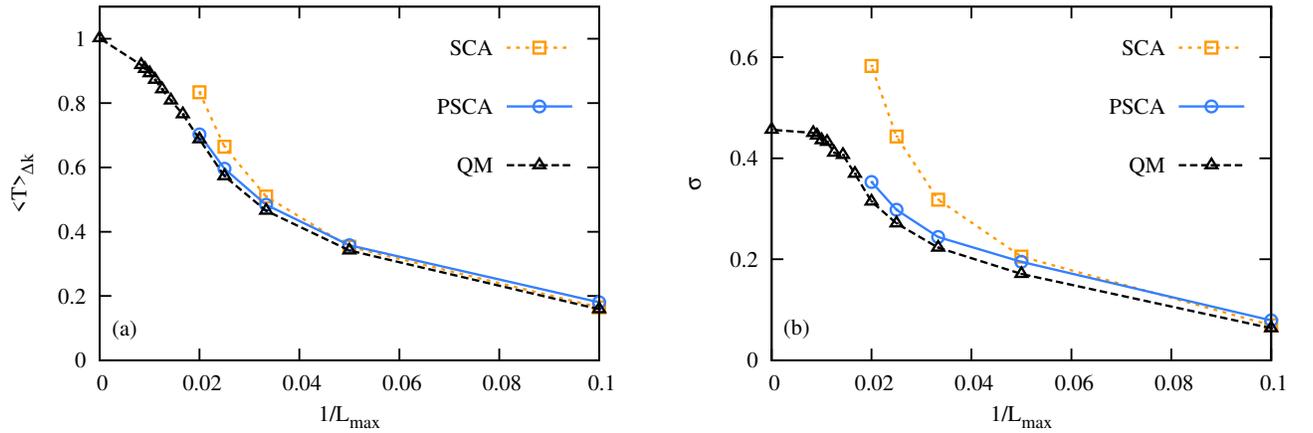}
	\caption{(Color online) Comparison between the PSCA, the SCA, and quantum mechanics (QM) for (a) the averaged transmission $\langle T\rangle_{\Delta k}$ as a function of the inverse cut-off length $1/L_{\rm max}$ and (b) the averaged variance of the conductance fluctuations $\sigma$. The average is performed in the interval $k\in[2.2,2.8]\pi/d$. The data within the PSCA are calculated in $\rm{3^{rd}}$ order for lengths $L_{\rm max}=10-30$, in $\rm{4^{th}}$ order for $L_{\rm max}=40$ and $\rm{5^{th}}$ order for $L_{\rm max}=50$. The QM results extend to the exact value ($L_{\rm max}=\infty$).}
\label{fig:mean_cond_var}
\end{figure*}
Conductance fluctuations (CF) have been identified as a direct manifestation of phase coherent transport (see e.g.~Refs.~\onlinecite{AkkMon06,Dat95,BarJalSto93,RahBro06,TwoTajBee04,JacSuk04,KhaEfe08,WeiRotBur05} and references therein). The strong fluctuations of the conductance as a function of, e.g., energy originate from path interference and thus give evidence for the wave nature of electrons in quantum dots. The CF offer one of the most stringent testing grounds for a semiclassical theory, when good agreement on the level of each individual $S$-matrix element is required.\\
In the following we compare the results for CF within the PSCA with the GTD-UTD, the SCA and the quantum mechanical calculations as a function of $k$ at vanishing magnetic fields $B=0$. Results for $B\ne 0$ averaged over $k$ can be found in Sec.~\ref{sec:wl}.\\
The semiclassical and quantum mechanical results both display strong fluctuations of the conductance [i.e., the total
transmission $T(k,B=0)$]. Their amplitude is, however, extremely sensitive to any deficiencies in the semiclassical approximations (inappropriate weighting of paths, missing paths). We emphasize that the comparison of Fig.~\ref{fig:kdep} is on a fully differential level. No energy or ensemble average is involved. Unsurprisingly, the agreement between the SCA and the quantum data is poor and on a level of qualitative agreement at best [Fig.~\ref{fig:kdep} (a) and (b)]. 
The functional dependence of $T(k,B=0)$ and $R(k,B=0)$ seems only weakly related to the quantum mechanical prediction. By ensemble averaging (e.g., over a suitable $k$ interval) these discrepancies would be (partially) removed (or masked). To correctly 
reproduce the fluctuations in $T(k,B=0)=\sum_{m,n}^M |t_{mn}(k,B=0)|^2$, all individual mode-to-mode amplitudes $t_{m n}$ have to be accurate. Obviously, the contribution of pseudo-paths included in the PSCA but missing in the SCA significantly improves the agreement with the quantum conductance fluctuations [Fig.~\ref{fig:kdep} (c) and (d)].\\
Even for averaged quantities, such as the averaged conductance $\langle T\rangle_{\Delta k}$ and the variance of the conductance fluctuations $\sigma=\sqrt{\langle T^2 \rangle_{\Delta k}-\langle T\rangle_{\Delta k}^2}$, the deficiencies of the SCA are still visible, in particular at larger cut-off lengths $L_{\rm max}$. Both $\langle T\rangle_{\Delta k}$ and $\sigma$ are overestimated (Fig.~\ref{fig:mean_cond_var}) as compared to quantum mechanics which can be attributed, in part, to the lack of correlation among purely classical paths.\\
In contrast, the PSCA shows very good agreement with quantum mechanics for both the averaged conductance as well as for the variance. For averaged quantities (Fig.~\ref{fig:mean_cond_var}) the inclusion of up to 4$^{\rm th}$ order diffractive scattering for $L_{\rm max}$ = 40 is sufficient while for fully differential quantities (Fig.~\ref{fig:kdep}) $5^{\rm th}$ order corrections for $L_{\rm max}$ = 40 still improve the agreement.\\
We point out that for the present system, taking into account paths up to a length of $L_{\rm max}=50$ which corresponds to $16$ radial traversals through the billiard, one reaches $70\%$ of the unitarity level within both the truncated quantum mechanics and the PSCA. The essential contributions to  CF, WL and shotnoise (see the following sections) is thus rooted in this subset of relatively short paths.\\
%
%---------------------------------------------------------------------------
\subsection{Weak localization}  \label{sec:wl}
%---------------------------------------------------------------------------
%
\begin{figure}[t]
	\centering
		\includegraphics[angle=-90,width=8.2cm]{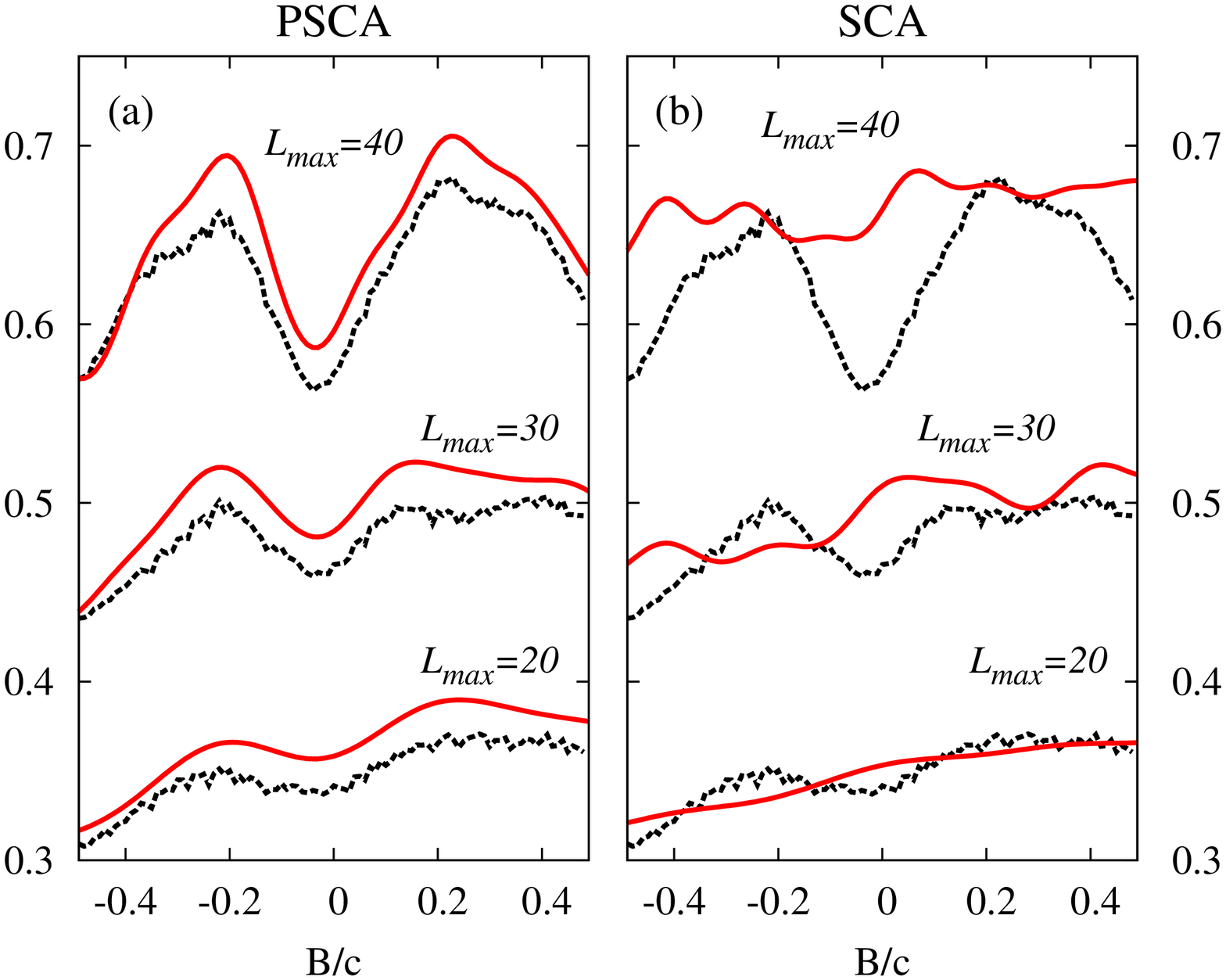}
	\caption{(Color online) Comparison between truncated quantum mechanics, the SCA, and the PSCA for the weak localization dip in transmission for different $L_{\rm max}$. (a) PSCA using the GTD-UTD for all diffraction coefficients (red solid line). (b) Standard SCA (red solid line) (using the GTD-UTD for the diffraction coefficients $c_m(\theta,k,d)$ for entering and exiting the circular cavity.) Black dashed lines: quantum results.}
\label{fig:wllength}
\end{figure}
\begin{figure}[t]
	\centering
		\includegraphics[angle=-90,width=8.2cm]{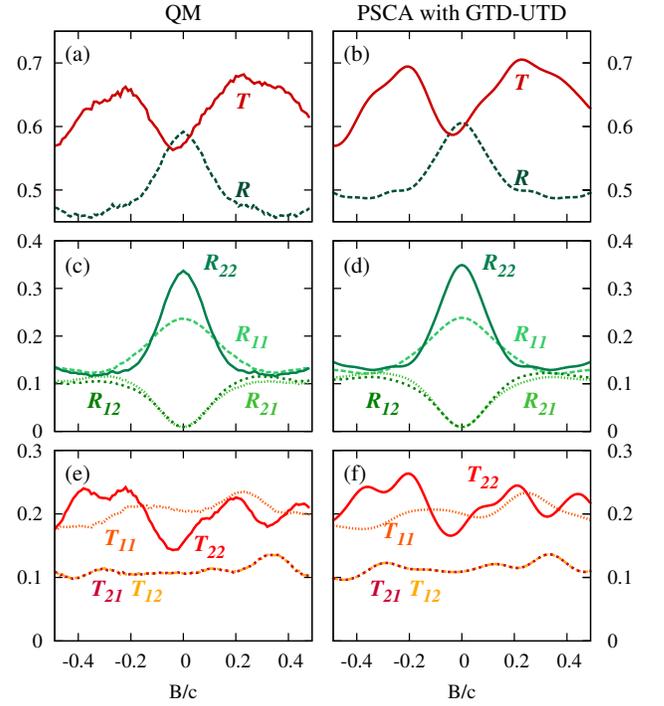}
	\caption{(Color online) Weak localization for $L_{\rm max}=40$ as produced by quantum mechanics: left column (a),(c),(e), and PSCA with the GTD-UTD: right column (b), (d), (f).}
\label{fig:wlpsca_gtdutd}
\end{figure}
\begin{figure}[t]
	\centering
		\includegraphics[angle=-90,width=8.2cm]{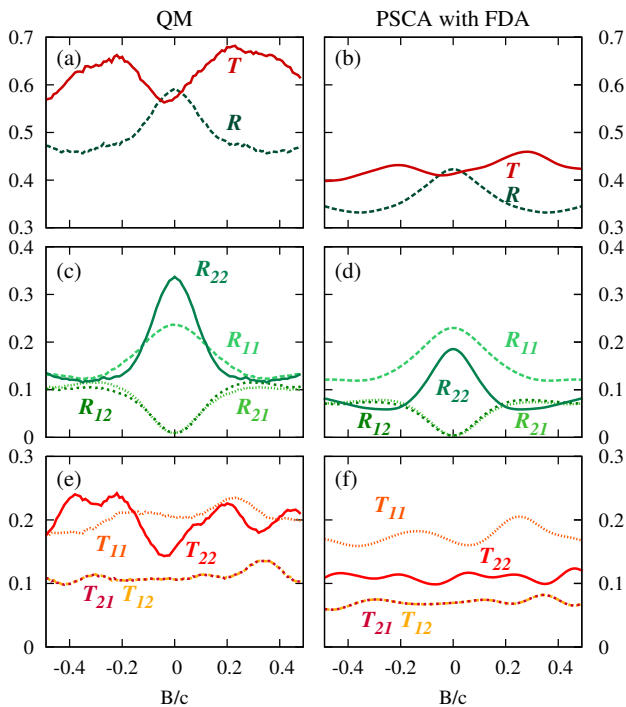}
	\caption{(Color online) Weak localization for $L_{\rm max}=40$. Left column  (a),(c),(e): quantum calculations, right column (b),(d),(f): PSCA with diffraction coefficients from FDA.}
\label{fig:wlpsca_fda}
\end{figure}
Weak localization (WL) is a well-known quantum correction to the classical diagonal approximation (see e.g.~ Refs.~\onlinecite{AkkMon06,Dat95,BarJalSto93,BreStaWirRotBur08,RicSie02,RahBro06,JacWhi06,RahBro05,ChaBarPfeWes94,HarWelMulRicSch08,LarMarTigCam04,RobSchOroFal08,KopSchRot08,GopRotSch06} and references therein) where the later corresponds to the restriction to terms $q = q'$ in the double sum over paths when calculating $|t_{nm}^{\rm SCA}|^2$ from Eq.~(\ref{fig:laspec}). Off-diagonal terms $q \neq q'$ give rise to quantum interferences which manifest themselves as an increase of the averaged total reflection 
$\langle R(B) \rangle _{\Delta k}=\sum_{m,n} \langle R_{nm}(B) \rangle _{\Delta k}$ at $B=0$ in form 
of a pronounced peak. Correspondingly, the averaged total transmission 
$\langle T(B) \rangle _{\Delta k}=\sum_{m,n} \langle T_{nm}(B) \rangle _{\Delta k}$ features a dip which is an immediate consequence of unitarity. For the investigation of the weak localization dip (peak) we employ an average over a small window $\Delta k=[2.2-2.8]\pi/d$ of the $k$ dependence of 
the probabilities: $\langle T_{nm}(B) \rangle _{\Delta k}=\int_{\Delta k} dk |t_{nm}(k,B)|^2$, 
$\langle R_{nm}(B) \rangle _{\Delta k}=\int_{\Delta k} dk |r_{nm}(k,B)|^2$.\\
Since semiclassical theories are, by construction, not necessarily unitary at a given level of approximation, the anti-correlated peak-dip structure near $B = 0$ provides a sensitive test for semiclassical approximations. It has been shown that the quantum anti-correlation between reflection and transmission $(\langle R(B)\rangle_{\Delta k}-\langle R(B=0)\rangle_{\Delta k})=-(\langle T(B)\rangle_{\Delta k}-\langle T(B=0)\rangle_{\Delta k})$ requires a correlation of transmitted and reflected paths.\cite{BreStaWirRotBur08} This correlation is absent in the SCA such that no transmission dip is reproduced [see Fig.~\ref{fig:wllength} (b)].\\
The role of diffraction in a quantum billiard manifests itself by a very good agreement of the PSCA with quantum mechanical results [see Fig.~\ref{fig:wllength} (a)]. Tests for WL as a function of the cut-off length $L_{\rm max}$ of the paths as well as of the order of diffractive scattering included show that convergence toward the (truncated) quantum result is reached for $L_{\rm max}$ = 40 when diffraction up to fourth order is included. We note an improvement compared to previous third-order calculations\cite{BreStaWirRotBur08} especially due to the inclusion of the (small) real part of the diffraction coefficient $v(\theta',\theta,k,d)$.\\
It is instructive to analyze the build-up of the weak localization peak in reflection and of the dip in transmission from individual $S$-matrix elements. In reflection only the diagonal elements $R_{11}(B)$ and $R_{22}(B)$ show a peak while the off-diagonal 
elements $R_{12}(B)$ and $R_{21}(B)$ exhibit a dip [Fig.~\ref{fig:wlpsca_gtdutd} (c), (d)]. This is due to the fact that time-reversal symmetric paths contributing to $R_{nm}(B)$  with different parity of modes $m$ and $n$ acquire an additional phase-shift of $\pi$. As an example consider $R_{21}(B)$, entering in mode $m=1$ and exiting in mode $m=2$ produces a phase shift of $\pi$. Thus the time-reversal symmetric paths interfere destructively.\\
In transmission the major contribution to a dip originates from $T_{22}(B)$ in the chosen energy window of two open modes [Fig.~\ref{fig:wlpsca_gtdutd} (e), (f)]. Note that $T_{21}(B)=T_{12}(B)$ because of the 
Onsager relation $t_{nm}^{(2,1)}(k,B)=t_{mn}^{(1,2)}(k,-B)$ and the symmetry of the circular billiard 
which ensures that $t_{mn}^{(1,2)}(k,-B)=t_{mn}^{(2,1)}(k,B)$. Due to time-reversal symmetry, $R(B)=R(-B)$ giving rise to a symmetric peak in $R$ as a function of $B$ (Fig.~\ref{fig:wlpsca_gtdutd}). For unitary transport, $T(B)=1-R(B)$, which implies a symmetric dip in transmission. For the truncated quantum mechanics and semiclassics where long paths are omitted unitarity is not preserved and $T(B)$ is not exactly symmetric but features a slight shift of the minimum toward $B<0$.\\
Overall, the agreement between the PSCA and the full quantum calculation (truncated at the same pathlength) is remarkable (Fig.~\ref{fig:wlpsca_gtdutd}). The residual minor deviations are mainly ascribed to deficiencies in the diffraction coefficients for which we use an analytical far-field approximation (Sec.~\ref{sec:diff} and appendix \ref{appA} and \ref{appB}). As, e.g., the diffraction coefficients $c_m(\theta,k,d)$ enter $R_{nm}$, $T_{nm}$ to the fourth power, a small deficiency in $c_m(\theta,k,d)$ can have a sizeable effect on $R_{nm}$, $T_{nm}$. The offset of the total reflection $R$ and transmission $T$ within PSCA [Fig.~\ref{fig:wlpsca_gtdutd} (b)] compared to the quantum mechanical result [Fig.~\ref{fig:wlpsca_gtdutd} (a)] mainly originates from imperfections in $R_{22}$ and $T_{22}$ [compare Fig.~\ref{fig:wlpsca_gtdutd} (c), (d), and (e), (f)]. The imperfection could be cured by including a correction factor of $\approx0.97-0.99$ in $c_{2}(\theta,k,d)$.\\
To demonstrate how sensitively transport properties depend on the weights of the contributing paths we show the results for weak localization calculated within PSCA but now using the FDA instead of the GTD-UTD for the diffraction coefficients (Fig.~\ref{fig:wlpsca_fda}). (We have used $\rm{3^{rd}}$ order of the PSCA here, since the deviation between $\rm{3^{rd}}$ and $\rm{4^{th}}$ order is much smaller than the errors due to the simpler diffraction theory.) It is striking that especially both $R_{22}$ and $T_{22}$ seem to be underestimated which is due to deficiencies in $c_2(\theta,k,d)$ within the FDA. It is worthwhile pointing out that the FDA diffraction could not be ``repaired'' by a correction factor, as the errors in $R_{22}$ and $T_{22}$ are different. 
In other words, transmission is more ``sensitive'' to a correct implementation of pseudo-paths than reflection. Of particular importance are paths which change the rotational direction.\cite{BreStaWirRotBur08} Since the FDA does not give sufficiently weight to this class of pseudo-paths (see Fig.~\ref{fig:diff_qm}) the transmission dip cannot be well reproduced.
%------------------------------------------------------------------------------------------------------------------------------
\subsection{Shot noise} \label{sec:shotnoise} 
%------------------------------------------------------------------------------------------------------------------------------
%
\begin{figure}[t]
	\centering
		\includegraphics[angle=-90,width=8.2cm]{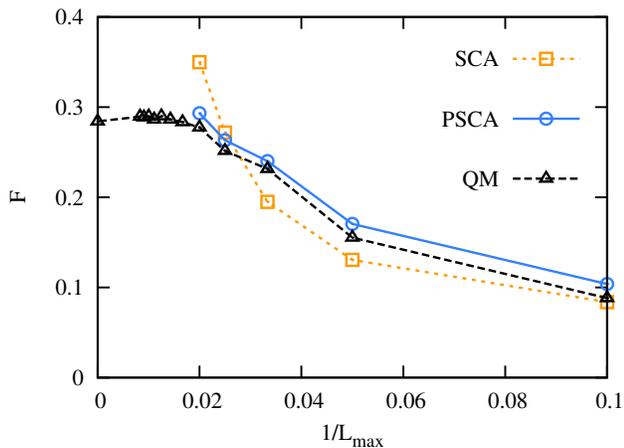}
	\caption{(Color online) The $k$-averaged Fano factor $F$ as a function of the inverse cut-off length $L_{\rm max}$. The average is performed over the interval $k\in[2.2,2.8]\pi/d$. Comparison between PSCA with the GTD-UTD, the SCA, and quantum mechanics (QM). ($L_{\rm max}=\infty$ corresponds to the exact result.)}
\label{fig:fano}
\end{figure}
Another quantity characteristic for quantum transport is the quantum shot noise power of the current (see e.g.~Refs.~\onlinecite{BlaBut00,Dat95,RahBro06,JacSuk04,AigRotBur05,RotAigBur07,KhoSavSom09,Nov07,BraHeuMulHaa06,TwoTajSchBee03,LewMucCas08,TwoTraTitRycBee06} and references therein). At zero temperature ($T = 0$), the time-dependent
current noise is due to the granularity of the electron charge and carries
information about the wave vs.~particle nature of charge transport.
The Fano factor $F$ measures the amount by which the noise in phase coherent
transport is suppressed relative to the Poissonian value of uncorrelated
classical electrons.
Within the Landauer-B\"uttiker picture, $F$ can be expressed
as\cite{BlaBut00}
\begin{equation}\label{eq:fanoshot}
F=\frac{\langle Tr(t^\dagger tr^\dagger r)\rangle_{\Delta k}}
{\langle Tr(t^\dagger t)\rangle_{\Delta k}}=
\frac{\langle \sum_n\tau_n\,\eta_n\rangle_{\Delta k}}{\langle \sum_n \tau_n\rangle_{\Delta k}},
\end{equation}
with $\tau_n$, $\eta_n$ being the eigenvalues of the Hermitian matrices
$t^\dagger t$ and $r^\dagger r$, respectively.\\
Calculating the shot noise power from non-unitary scattering 
matrices is obviously a delicate matter, as replacement of $r^t r$ by $1 - t^t t$ leads, unlike for unitary descriptions, to different results. Furthermore, 
such a replacement may result in 
negative and thus unphysical values for the shot noise power, as
non-unitary scattering matrices 
allow for the possibility of having $\tau_n>1$ such that $(1-\tau_n)<0$ 
(for very high mode numbers as, e.g., in Ref.~\onlinecite{RahBro06} such a 
situation may, however, be unlikely). By using
Eq.~(\ref{eq:fanoshot}), such difficulties can be avoided as both the transmission
and reflection eigenvalues $\tau_n,\,\eta_n$ are, by construction, real and 
positive not only for the truncated quantum calculation but also for
the PSCA and the standard SCA. The standard SCA result for $F$
strongly deviates from the quantum mechanical data (see Fig.~\ref{fig:fano}). For small cut-off 
lengths $L_{\rm max}$, the value of $F$ is smaller but increases more rapidly than the quantum mechanical result with $L_{\rm max}$. The PSCA yields very good
agreement with the quantum mechanical result
for the shot noise Fano factor $F$. Note that for the largest cut-off length $L_{\rm max}=50$ the Fano factor $F$ is already converged to its asymptotic value $F\approx 0.28$ suggesting that long paths do not play a significant role for $F$. This result agrees with the finding\cite{AigRotBur05,RotAigBur07} that the shot noise power is of similar magnitude for regular and chaotic billiards as differences in the dynamics are most strongly felt by very long paths.
%------------------------------------------------------------------------------------------------------------------------------
\section{Summary}
%------------------------------------------------------------------------------------------------------------------------------
We have presented a semiclassical theory which is able to {\it quantitatively} reproduce full quantum
results for scattering through microstructures with a specific geometry, in the present case a circular shaped billiard with leads oriented 90$^o$ degrees relative to each other. The present approach does not invoke the limit of large mode numbers, (where the de Broglie wavelength $\lambda$ is small relative to the lead width) but requires $\lambda$ to be small only on the scale of the linear dimension of the microstructure (the circle). This non-asymptotic semiclassical theory allows a direct comparison with quantum calculations as well with experiments on a system-specific level for individual $S$-matrix elements avoiding any ensemble averaging or fit parameters. 
This level of agreement allows us to perform detailed semiclassical investigations of quantum transport quantities such as the conductance fluctuations, the weak localization, and the shot noise. Our studies show unambiguously that for reproducing these quantities correctly, two major ingredients are, indeed, crucial: (1) the inclusion of 
``pseudo-paths'' in the semiclassical propagator which are diffractively backscattered from the interior side of the
cavity openings, (2) a sufficiently accurate description of the diffraction coefficients for the injection, 
ejection and back-reflection of particle flux at the cavity openings. We meet the latter requirement by developing a combined geometric and uniform theory of diffraction (GTD-UTD). 
Pseudo-paths are crucial for reproducing the conductance fluctuations in transport and lead to 
a reduction of its variance. Also for the weak-localization effect we find that
pseudo-paths are crucial, as no signature of weak localization appears 
in the transmission through the circular
cavity without their contribution (even when the advanced diffraction 
theory is employed for all truly classical paths). For the shot noise 
power we showed that a standard semiclassical calculation (without
pseudo-paths) gives sizeable discrepancies. The inclusion of pseudo-paths leads to agreement with the quantum mechanical result. We emphasize that the parameter regime in which we have identified the above effects of pseudo-paths coincides with the typical situation in quantum transport experiments. The latter usually feature only a few open lead modes $M$.\\ 
The present results raise several interesting questions for future investigations: the comparison between the semiclassical approximations (PSCA and SCA) and full quantum calculations were performed for truncated path sums up to a finite path length $L \leq L_{\rm max}$. The primary reason for the truncation was technical, as the number of diffractive pseudo-paths exponentially proliferates with $L \rightarrow \infty$ also for classically regular structures and exact path sums become prohibitively difficult to perform. There is, however, a second conceptual motivation. In the experiment, decoherence due to inelastic scattering limits phase-coherent transport to pathlengths $L \leq l_\phi$, where the phase-decoherence mean free path $l_\phi$ typically allows only a moderate number of traversals across the cavity. 
The latter restriction rules out that very long paths with $L>l_\phi$ contribute to quantum interference in the experiment,
a feature which is naturally incorporated by way of the cut-off length $L_{\rm max}$ in our semiclassical theory. Clearly, such long paths can still provide incoherent contributions. The present approach may thus contribute to a semiclassical unterstanding of decoherence effects in regular cavities.\cite{ChaBarPfeWes94}\\
In the present treatment of diffractive scattering, both internal diffraction at the open lead mouth giving rise to pseudo-paths as well as the coupling between leads and cavity was performed for sharp edges. The weight of diffractive contributions can be changed by ``rounding off'' the lead opening. An investigation of the dependence of the weak localization on the smoothness of the edges is currently underway.\footnote{T.~Dollinger \emph{et al.},~(unpublished).} The introduction of rounded corners has, however, another profound effect, apart from changing the weight of diffractive scattering: an open circle with rounded edges of the leads is no longer regular but features a mixed phase space. This raises the question as to the interplay between diffractive scatterings at the lead opening and chaotic scattering in the interior of the billiard. For generic chaotic systems a number of alternative semiclassical theories has been proposed to explain
universal features of quantum transport (see e.g.~Refs.~\onlinecite{RicSie02,BraHeuMulHaa06,HeuMulBraHaa06,RahBro05,RahBro06,BroRah06,JacWhi06} and references therein). These theories typically employ an ensemble average and rely on a $\hbar\to 0$ limit which makes them complementary to the present system-specific approach for finite $\hbar$. Bridging the
gap between these two frameworks would be of great interest. One key ingredient would be to clarify the interplay between 
the diffraction-based pseudo-paths and the chaos-based correlated classical path pairs (Richter-Sieber orbits\cite{RicSie02}).
The relative weight for characteristic quantum transport effects carried by these two classes of paths when both are present (as in a chaotic billiard with sharp edges) remains an open question.
\section*{Acknowledgments}
We thank Piet Brouwer, Tobias Dollinger and Klaus Richter for helpful discussions. This work was supported by the Austrian FWF under Grants No.~FWF-SFB016 ``ADLIS'' and No.~17359, and the FWF doctoral program ``Complex Quantum Systems''. L.W. acknowledges support by the PHC ``Amadeus'' of the French Ministry of Foreign and European Affairs. 

%---------------------------------------------------------------------------------------------------------------------------------
\appendix
\section{Geometric and uniform theory of diffraction
for the coupling of quantum leads to a billiard cavity}
%----------------------------------------------------------------
In a quantum billiard, the electron propagation in the leads
is determined by a quantum mechanical wave, while the propagation
inside the ballistic cavity can be described semiclassically, i.e., by propagation
along classical trajectories with a quantum phase.
The strong scattering effects that occur especially in the low mode regime at the orifices
have been described in past works on transport through 
open billiards by the Kirchhoff diffraction approximation (KDA)
\cite{SchAlfDel96} or on the level of the Fraunhofer
diffraction approximation (FDA).\cite{WirTanBur97,WirTanBur99,WirStaRotBur03,StaRotBurWir05}
In the high-mode regime diffraction has been neglected
altogether.\cite{BarJalSto93,RicSie02,BraHeuMulHaa06,HeuMulBraHaa06,RahBro05,RahBro06,BroRah06,JacWhi06,PicJal99,arg96b} Both the KDA and the FDA perform about equally well
(see, e.g., Fig.~\ref{fig:diff_qm}). For the identification of pseudo-paths in the Fourier spectra of the conductance fluctuations,
the KDA and the FDA have been sufficiently precise. However, in order to
recover unitarity of the semiclassical $S$-matrix and weak-localization in transmission, a more precise diffraction theory must be implemented. For multiple-scattering
paths, higher
order products of the diffraction weights occur and even small
errors are rapidly amplified. We have therefore implemented a combination
of the geometric theory of diffraction (GTD) \cite{Kel62} and the uniform 
theory of diffraction (UTD) \cite{KouPat74,SiePavSch97} for the diffraction coefficients 
in open billiards referred to in the following as the GTD-UTD. Both theories have been 
previously applied separately to the 
calculation of higher-order scattering corrections to Gutzwiller's 
trace formula in closed quantum-billiards.\cite{SiePavSch97}
We discuss in section \ref{appA} diffractive backscattering at
a semi-infinite half-plane with an orifice and in section \ref{appB} diffraction during propagation from a lead into a 
semi-infinite plane.\\
%----------------------------------------------------------------
\subsection{The GTD-UTD for backscattering into the cavity} \label{appA}
%----------------------------------------------------------------
%
\begin{figure}[t]
	\centering
		\includegraphics[width=5cm]{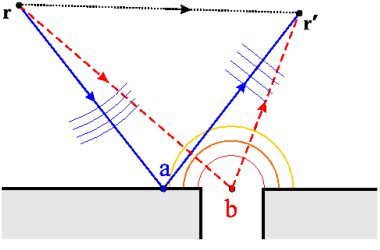}
	\caption{(Color online) Propagation between two points $\vec{r}$ and $\vec{r}\,'$ in a semi-infinite plane
in the presence of an open (lead). Contributions from the direct path (black dotted line),
from specular reflection (blue solid line) ,
and from scattering at the orifice (red dashed line) are depicted.}
\label{fig:half_inf}
\end{figure}
\begin{figure}[t]
	\centering
		\includegraphics[width=8.2cm]{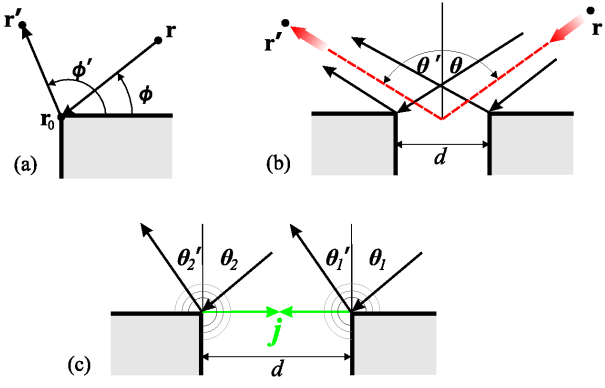}
	\caption{(Color online) (a) Diffraction at a wedge. (b) Diffraction at a lead described as diffraction at two wedges. The reference 
path is marked by a red dashed line. The angle is counted positive for a path lying on the left of the vertical axis, for a path on the right the angle is negative. (In the present example $\theta$ is negative and $\theta'$ is positive). The points $\vec{r}$ and $\vec{r}\,'$ are assumed to be in the far field region. (c) Higher-order corrections enter via paths scattered between the wedges.}
\label{fig:diff_wedge}
\end{figure}
\begin{figure}[t]
	\centering	\includegraphics[angle=-90,width=8.2cm]{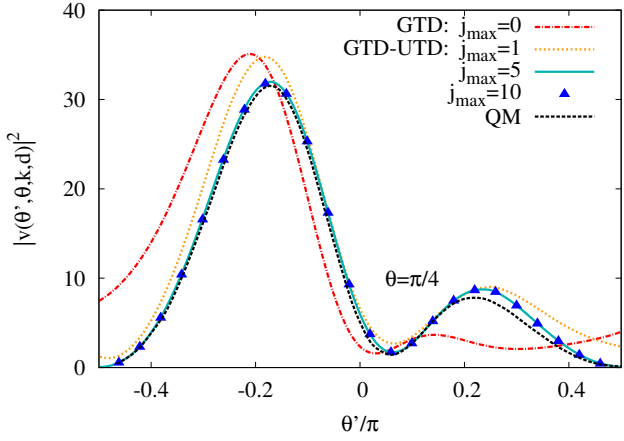}
	\caption{(Color online) Absolute square of the diffraction coefficient $|v^{\rm{GTD}}(\theta',\theta,k,d)|^2$ (or $v^{\rm GTD-UTD}$ with $j_{\rm max}=0$) compared with $|v^{\rm{GTD-UTD}}(\theta',\theta,k,d)|^2$ for $j_{\rm max}=1$, $j_{\rm max}=5$ and $j_{\rm max}=10$ (for the definition of $j_{\rm max}$ see text), $k=2.5\pi/d$. The quantum mechanical (QM) result is taken from Ref.~\onlinecite{SchAlfDel96}.}
\label{fig:diff_convergence}
\end{figure}

The propagation between two points in a semi-infinite plane
with a connected lead is depicted in Fig.~\ref{fig:half_inf}.
In the absence of a lead, the propagation between
two points $\vec{r}$ and $\vec{r}\,'$ in a semi-infinite plane
is described by a sum of two Green's functions, 
$G(\vec{r}\,',\vec{r},k) = G^{\rm dir}(\vec{r}\,',\vec{r},k)
+ G^{\rm refl}(\vec{r}\,',\vec{r},k)$.
The first contribution corresponds to the direct path
from $\vec{r}$ to $\vec{r}\,'$ (black dotted line) and
is in the semiclassical approximation [Eq.~(\ref{eq:gfree})] given by
\begin{eqnarray}
G^{\rm dir}(\vec{r}\,',\vec{r},k) & = & \frac{e^{ikL(\vec{r}\,',\vec{r})-i3\pi/4-i\mu\pi/2}}
{\sqrt{2\pi k L(\vec{r}\,',\vec{r}\,)}} \nonumber \\
& =: & G^{\rm SCA}(\vec{r}\,',\vec{r},k),
\end{eqnarray}
where $L(\vec{r}\,',\vec{r}$) is the distance between the two points and $\mu$ is the Maslov index ($\mu=0$ in this case).
The second term corresponds to a propagation
via a classical, specularly reflected path (blue solid line):
\begin{equation}
G^{\rm refl}(\vec{r}\,',\vec{r},k) = \frac{e^{ik\left[L(\vec{r}\,',\vec{a})+L(\vec{a},\vec{r}\,)\right]-
i3\pi/4-i\pi}}
{\sqrt{2\pi k \left[L(\vec{r}\,',\vec{a})+L(\vec{a},\vec{r}\,)\right]}}.
\end{equation}
The orifice gives rise to an additional scattered wave for which we assume in the far-field limit ($|\vec{k}\vec{r}\,| \gg 1$
and $|\vec{k}\vec{r}\,'| \gg 1$) a cylindrical wave emanating
from the center of the orifice (point $\vec{b}$). Invoking far-field approximations is at the heart of semiclassical diffraction theories, the validity of which need testing on a case-by-case basis (see below).\\
The diffraction contribution to the Green's function
(red dashed line in Fig.~\ref{fig:half_inf}) is
\begin{eqnarray}
G^{\rm PSCA}(\vec{r}\,',\vec{r},k)   & = & 
\frac{e^{ikL(\vec{r}\,',\vec{b})-i3\pi/4}}
{\sqrt{2\pi k L(\vec{r}\,',\vec{b})}}v(\theta',\theta,k,d) \times
\nonumber \\
& & 
\frac{e^{ikL(\vec{b},\vec{r})-i3\pi/4}}
{\sqrt{2\pi k L(\vec{b},\vec{r}\,)}} \nonumber \\
& = & G^{\rm SCA}(\vec{r}\,',\vec{b},k) v(\theta',\theta,k,d) \times
\nonumber \\
& & G^{\rm SCA}(\vec{b},\vec{r},k).
\label{pscadef}
\end{eqnarray}
We refer to this term as the (first-order) pseudo-path semiclassical contribution, because it corresponds to a classically forbidden path.
We note the different scaling of $G^{\rm PSCA}$ and
$G^{\rm refl}$ with $k$. The amplitude of $G^{\rm refl}$ scales as $1/\sqrt{k(L_1+L_2)}$ while the diffractive contribution scales as $1/\sqrt{k^2L_1L_2}$. At large distances (far-field) and/or large $k$, the geometric reflection amplitude dominates over the diffraction amplitude, as expected.\\
The diffraction coefficient $v(\theta',\theta,k,d)$ as a function
of the incoming and outgoing angles has been calculated in the
past in the Kirchhoff (KDA) and Fraunhofer diffraction approximations (FDA).
Both approaches have in common that the amplitude of the
cylindrical diffractive wave emanating from the orifice is determined from 
an integration over the orifice using as a source the amplitude and the phase of the unperturbed 
incoming wave in the absence of the boundary. In other words, 
they are implementations of Huygens' principle according to which each point
of the lead opening is the source of an outgoing circular wave.
For details, we refer the reader to Ref.~\onlinecite{SchAlfDel96}
(for the KDA) and to Ref.~\onlinecite{WirTanBur97,WirStaRotBur03} (for the FDA).\\
Keller's geometric theory of diffraction (GTD)\cite{Kel62} has a different 
point of departure: it originates from the far-field approximation of the exact Sommerfeld's solution for the wave scattering at a wedge \cite{Som50} [see Fig.~\ref{fig:diff_wedge} (a)]. In the following we discuss the application of the GTD and its refinement within the framework of the uniform theory of diffraction UTD \cite{KouPat74} to the problem of the diffractive scattering at the lead mouth [Figs.~\ref{fig:diff_wedge} (b) and (c)]. Starting point of our determination of the diffraction coefficient $v (\theta', \theta, k,d)$ is the decomposition of the orifice into two wedges with an inner angle
of $\pi/2$ and outer angle of $3\pi/2$ [see Fig.~\ref{fig:diff_wedge} (a)].
In the far-field limit, the incident wave can be regarded asymptotically as a plane-wave.
The GTD describes the scattering of a plane
wave at an infinitely sharp wedge \cite{Kel62}
[Fig.~\ref{fig:diff_wedge} (a)]. 
The total wavefunction is 
a sum of the incoming plane wave, the reflected plane wave and
an outgoing cylindrical diffracted wave, emanating from the edge.
There are two discontinuities: at the
shadow boundary [$\phi'-\phi =\pi$, for the definition of the angles
see Fig.~\ref{fig:diff_wedge} (a)] and the boundary of geometric 
reflection ($\phi'+\phi =\pi$). Outside a small region around
these two angles, the diffracted wave can
be described by an outgoing cylindrical wave modulated by a 
smooth diffraction coefficient. The diffractive part of the Green's
function between the two points $\vec{r}$ and $\vec{r}\,'$ in
Fig.~\ref{fig:diff_wedge} (a) can thus be approximated as:\footnote{The factor $1/2$ (atomic units) in Eq.~(\ref{Gpsca}) originates from $\hbar^2/2m$ (SI units) 
and assures that the Green's function is correctly normalized 
[such that $(\frac{\hbar^2k^2}{2m}-\hat{H}_{\vec{r}\,'})G(\vec{r}\,',
\vec{r},k)=\delta(\vec{r}\,'-\vec{r})$]. We use the same normalization 
as in Ref.~\onlinecite{BraBha03}, section 7.5.4.}
\begin{equation} 
G^{\rm PSCA}(\vec{r}\,',\vec{r},k) = 
G^{\rm SCA}(\vec{r}\,',\vec{r}_0,k) 
\frac{1}{2}
D(\phi',\phi),
G^{\rm SCA}(\vec{r}_0,\vec{r},k),
\label{Gpsca}
\end{equation}
with
\begin{eqnarray}
D(\phi',\phi)&=& -2\frac{\sin{\pi/N}}{N} \Bigg[\frac{1}{\cos{\frac{\pi}{N}}-\cos{\frac{\phi'-\phi}{N}}}\nonumber \\ &-&
\frac{1}{\cos{\frac{\pi}{N}}-\cos{\frac{\phi'+\phi}{N}}}\Bigg],
\label{eq:D}
\end{eqnarray}
where $N=3/2$ is the exterior angle (in units of $\pi$) of a perpendicular 
wedge and $\vec{r}_0$ is the position of the corner of the wedge 
[Fig.~\ref{fig:diff_wedge} (a)].\\
We now consider the lead opening as being composed of two wedges
[Fig.~\ref{fig:diff_wedge} (b)]. The obvious conceptual difficulty lies in the fact that the two wedges are, in general, not in the far-field limit $(k d \gg 1)$ of each other. With this caveat in mind, diffraction at the lead can be considered within the GTD to result from the interference of two paths that are diffracted at the corners of the two wedges limiting the lead. Summing up the diffraction weights and phases from the two paths, we obtain the Green's functions of the reference pseudo-path as in Eq.~(\ref{pscadef}) but with the GTD 
reflection coefficient 
\begin{eqnarray}
v^{\rm{GTD}}(\theta',\theta,k,d)&=& \frac{1}{2}D_L(\theta',\theta)e^{-ik\frac{d}{2}(\sin\theta'+\sin\theta)}\nonumber \\ &+&
\frac{1}{2}D_R(\theta',\theta)e^{+ik\frac{d}{2}(\sin\theta'+\sin\theta)}. \nonumber \\
\label{eq:vgtd}
\end{eqnarray}
In Eq.~(\ref{eq:vgtd}) we have neglected the difference of the incoming angles at the left and at the right wedge, respectively,
i.e., we set $\theta_1=\theta_2=\theta$ [Fig.~\ref{fig:diff_wedge} (c)].
Likewise, for the outgoing angles, we set $\theta'_1=\theta'_2=\theta'$.
The phase differences of the right/left pseudo-path with respect to 
the reference path emanating from the center of the orifice can then be written in linear approximation with respect to the transverse lead coordinate as 
$k\Delta L = \pm ik\frac{d}{2}(\sin\theta'+\sin\theta)$.
The coefficients in Eq.~(\ref{eq:vgtd})
are defined as
\begin{eqnarray}
D_L(\theta',\theta) & = & D(\pi/2-\theta',\pi/2-\theta) \nonumber \\
D_R(\theta',\theta) & = & D(\pi/2+\theta',\pi/2+\theta).
\label{eq:gtdampl}
\end{eqnarray}
We note that both $D_L$ and $D_R$ have a singularity at the reflection boundary
for $\theta'=-\theta$. However, in the sum of the two terms, the
singularities cancel out and the resulting backscattering
amplitude is perfectly smooth (see Fig.~\ref{fig:diff_convergence},
red chain dotted line).\\
The comparison of the GTD with the exact quantum calculation for the diffraction coefficient, $|v (\theta, \theta', k, d)|^2$ (Fig.~\ref{fig:diff_convergence}), reveals sizeable deviations. Two deficiencies are noteworthy: the almost complete missing of the back-reflection peak and the failure at grazing angles $\theta', \theta \rightarrow \neq \pi/2$. The diffraction coefficient should approach zero in this limit but, instead, converges toward a finite value 
(Fig.~\ref{fig:diff_convergence}). This behavior originates from treating
the diffraction at the lead as two independent local phenomena of diffraction at two separate wedges.\\
The diffraction at a lead can be treated within the UTD by using a double-wedge diffraction coefficient (see Ref.~\onlinecite{Sch88,SchLue91} and a more recent paper for arbitrary configurations of the wedges, Ref.~\onlinecite{Alb05}). The double-wedge diffraction coefficient cannot be separated into a sequence of single-wedge diffraction coefficients and contains rather involved mathematical expressions such that the beauty and structural simplicity of a semiclassical approach is lost. We present in the following an ansatz for double-wedge diffraction which assumes a separation of the diffraction process into a sequence of diffraction events and we verify its validity by comparing with quantum mechanical results from Ref.~\onlinecite{SchAlfDel96}. We show that the shortcomings of Eq.~\ref{eq:vgtd} can be (to a large extent) remedied by taking diffractive paths of higher order into account, i.e. paths 
that pass between the two wedges once or several times. This drastically improves the agreement with the quantum mechanical result. (We point to the conceptual similarity of our approach to the treatment of double-wedge diffraction illuminated by transition region fields by a sum over higher-order diffracted fields.\cite{Hol96})
In a first step, we include paths that scatter once between the two
edges [see green line Fig.~\ref{fig:diff_wedge} (c)].
There are two such paths. One approaches the
right wedge at angle $\theta$. It is scattered into the angle $\pi/2$
(with respect to the surface normal), and at the left wedge, it is scattered
into the angle $\theta'$. The other path is scattered from the left
wedge to the right one with the same entrance and exit angles.
The weights of this pair of paths cannot be determined from the GTD 
diffraction coefficients [Eq.~(\ref{eq:D})]: The GTD diffraction coefficient fails
in the limit of $\phi \rightarrow 0$ and $\phi' \rightarrow \pi$
[definition of angles as in Fig.~\ref{fig:diff_wedge} (a)],
because this is in proximity of the shadow boundary, into which the horizontal paths are scattered. 
This problem can be overcome by invoking the uniform theory of diffraction (UTD).\\
Contrary to the GTD, the UTD is also valid on the zone boundaries.
The outgoing cylindrical wave is multiplied 
by a diffraction coefficient which depends not only
on the two angles $\phi'$ and $\phi$ but also on the distances
$r'$ and $r$, and on the wavenumber $k$,
\begin{eqnarray}
D^{\rm{UTD}}(\phi',\phi,r',r,k)=-\frac{e^{i\frac{\pi}{4}}}{N}&\times& \nonumber \\
\sum_{\sigma,\eta =\pm 1} \sigma\cot{\left(\frac{\pi+\eta(\phi'-\sigma\phi)}{2N}\right)} &\times&\nonumber \\
F\left(k\frac{rr'}{r+r'}a_\eta(\phi'-\sigma\phi)\right),
\label{eq:dutd}
\end{eqnarray}
where $a_{\pm}(\beta)=2\cos^2{\Big(\frac{2\pi Nn^{\pm}-\beta}{2}\Big)}$ and 
$n^{\pm}$ is the 
integer which most closely satisfies $2\pi Nn^{\pm}-\beta=\pm \pi$.
The function $F$ is defined as a generalized
Fresnel integral:
\begin{equation}
F(x)=-2i\sqrt{x}e^{-ix}\int_{\sqrt{x}}^{\infty} d\tau e^{i\tau^2} 
\end{equation}
and has the asymptotic form
\begin{equation}
F(x) = 1 - i\frac{1}{2x} - \frac{3}{4} \frac{1}{x^2} + \ldots.
\end{equation}
For $x \rightarrow \infty$, i.e., for large distances and outside the
transition zones, $F(x)=1$, and $D^{\rm{UTD}}(\phi',\phi,r',r,k)$ 
reduces to $D^{\rm{GTD}}(\phi',\phi)$.
Within the transition zone, the distance dependence of $D^{\rm{UTD}}$
leads to a deviation of the scattered wave from a purely cylindrical wave.
This is necessary to ensure the continuity of the total wave-function
at the zone boundaries. Furthermore, by using the UTD diffraction coefficient we partially take into account the fact that the two wedges are not in the far-field region with respect to each other.\\
Since the GTD 
fails for large (near grazing) angles, we opt for a piecewise construction: We combine GTD and UTD referred to in the following as GTD-UTD such that the diffraction of the path with the smaller (absolute value) of the angle is treated by the GTD and the path with the larger angle on the level of the UTD. Accordingly, to first order the GTD-UTD correction to the diffraction coefficient can be written as
\begin{eqnarray} 
& & v^{\rm{GTD-UTD,1}}(\theta',\theta,k,d) = \nonumber \\
& & \frac{1}{2}U_L(\theta',-\pi/2,d,k)\frac{e^{ikd(-\sin\theta'+\sin\theta)/2}}
{\sqrt{2\pi kd}} \frac{1}{2}D_R(+\pi/2,\theta) \nonumber \\
& + &
\frac{1}{2}U_R(\theta',+\pi/2,d,k)\frac{e^{ikd(+\sin\theta'-\sin\theta)/2}}
{\sqrt{2\pi kd}} \frac{1}{2}D_L(-\pi/2,\theta), \nonumber \\
\label{eq:vgtdutd1}
\end{eqnarray}
where [in analogy to Eq.~(\ref{eq:gtdampl})] we have defined the
UTD diffraction coefficients at the left (L) and right (R) wedge as
\begin{eqnarray}
U_L(\theta',\theta,r,k) & = &D^{\rm{UTD}}(\pi/2-\theta',\pi/2-\theta,r' \rightarrow \infty,r,k) \nonumber \\
U_R(\theta',\theta,r,k) & = &D^{\rm{UTD}}(\pi/2+\theta',\pi/2+\theta,r' \rightarrow \infty,r,k).\nonumber \\
\label{eq:Utd}
\end{eqnarray}
In Eq.~(\ref{eq:vgtdutd1}), the diffraction of the incoming path (with
angle $\theta$), is treated on the GTD level and the diffraction of the
outgoing path (angle $\theta'$) on the UTD level. This construction introduces a first-order discontinuity (``kink'') at $|\theta|=|\theta'|$. This kink is, however, negligible for small angles and visible only at large angles close to $\pi/2$, in other words the formula breaks down in the limit $\theta,\theta' \rightarrow \pi/2$, see, e.g., Fig.~\ref{fig:diff_qm} (c). (One could alternatively ignore the fact that the UTD is not multiplicative in the near-field and employ for both scattering events UTD, thus avoiding the kink. This ansatz, however, breaks down similarly for grazing incidence: the diffraction coefficient does not approach zero for the outgoing angle $\theta' \rightarrow \pi/2$. For small and medium angles this approach behaves equally well as the presented GTD-UTD approach.) At large incident and/or outgoing angles the diffraction coefficient is already strongly suppressed such that the resulting error is small.\\
The inclusion
of the first order GTD-UTD correction ($j_{\rm max}=1$, see Fig.~\ref{fig:diff_convergence}) already considerably improves the 
agreement with the quantum diffraction pattern. Higher order diffraction corrections include paths that are scattered several times between the wedges. This includes paths that are incident and 
backscattered at an angle $\pm \pi/2$ at the wedge. This is exactly on the reflection boundary. In this limit, Reiche has shown \cite{Rei11} that the diffraction pattern of a plane wave 
with unit amplitude incident on the wedge with an angle of $\pm \pi/2$ 
reduces to a reflected plane wave with amplitude $1/2$ and a cylindrical wave. 
The Green's function of a higher order path which scatters $j$ times between 
the wedges is, therefore, a product of the GTD diffraction coefficient, the 
UTD diffraction coefficient, and the Green's function for free propagation along 
the distance $jd$, acquiring a factor $1/2$ and a phase of $\pi$ for each 
reflection at a wedge. Summing the diffraction corrections up to order $j_{\rm max}$ we obtain
\begin{widetext}
\begin{eqnarray} 
v^{\rm{GTD-UTD}}(\theta',\theta,k,d)  =  v^{\rm{GTD}}(\theta',\theta,k,d)+ \nonumber \\
\frac{1}{4} \sum_{\rm{odd}:\ j=1}^{j_{\rm max}}  U_L(\theta',-\pi/2,jd,k)g_j(k)e^{i\Phi_{-+}}D_R(+\pi/2,\theta)  +   U_R(\theta',+\pi/2,jd,k)g_j(k)e^{i\Phi_{+-}}D_L(-\pi/2,\theta)+\nonumber \\
\frac{1}{4} \sum_{\rm{even}:\ j=1}^{j_{\rm max}} U_R(\theta',+\pi/2,jd,k)g_j(k)e^{i\Phi_{++}}D_R(+\pi/2,\theta)  + 
U_L(\theta',-\pi/2,jd,k)g_j(k)e^{i\Phi_{--}}D_L(-\pi/2,\theta), \nonumber \\
\label{eq:vgtdutd}
\end{eqnarray}
\end{widetext}
where 
\begin{equation}
g_j(k)=\frac{1}{\sqrt{2\pi kjd}}\frac{1}{2^{j-1}}e^{i(kjd+(j-1)\pi)}
\label{gkj}
\end{equation}
and
\begin{equation}
\Phi_{\pm \pm}=k\frac{d}{2}(\pm \sin{\theta'}\pm \sin{\theta}).
\end{equation}
Fig.~\ref{fig:diff_convergence} demonstrates that 
the diffraction coefficient $v(\theta',\theta,k,d)^{\rm GTD-UTD}$ is converged
for $j_{\rm max}=5$. Furthermore, the condition $v(\theta',\theta,k,d)\rightarrow 0$ 
for $\theta,\theta' \rightarrow \pi/2$ is fulfilled and the agreement
with the fully quantum mechanical backscattering weight is excellent.
The agreement deteriorates for increasing entrance angles $\theta$ but is still satisfying compared to the simple Fraunhofer diffraction approximation (FDA), 
(see Fig.~\ref{fig:diff_qm}). Fortunately, 
large angles do not play an important role because the overall diffraction
weight is very low. The GTD-UTD provides a remarkable 
compromise between simplicity and accurate representation of quantum mechanical 
results for diffraction at a lead attached to a semi-infinite half-plane.\\
%------------------------------------------------------------------------------------------------------------------------------
\subsection{The GTD-UTD for coupling
from lead modes into the cavity}\label{appB}
%------------------------------------------------------------------------------------------------------------------------------
\begin{figure}[t]
	\centering
		\includegraphics[width=6cm]{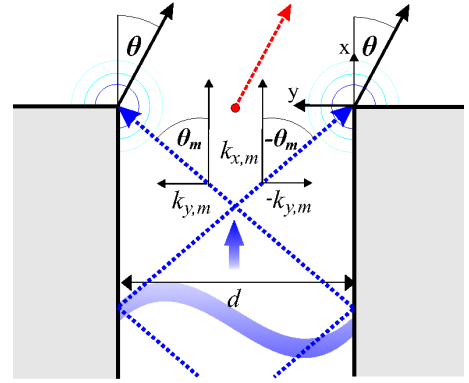}
	\caption{(Color online) A lead of width $d$ coupled to a half-infinite plane. An incoming wave in mode $m$ can be separated into two rays $\theta = \pm \theta_{m} =\pm\arcsin{m\pi/dk}$ which  diffractively scatter at the lead wedges. The reference path is denoted by a red dashed line.}
\label{fig:difflead}
\end{figure}
\begin{figure*}[t]
	\centering
		\includegraphics[angle=-90,width=17cm]{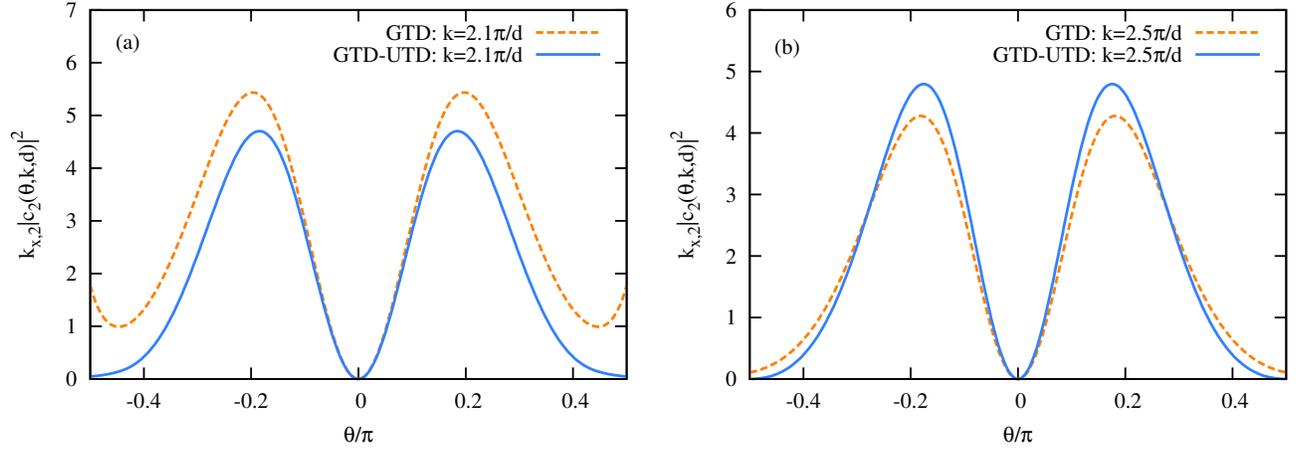}
	\caption{(Color online) The absolute square of the coupling coefficient $|c_m(\theta,k,d)|^2$ multiplied by $k_{x,m}$ for $m=2$ at two different wavenumbers (a) $k=2.1\pi/d$ and (b) $k=2.5\pi/d$ within the GTD and the GTD-UTD. Note the breakdown of the GTD for $\theta\rightarrow \pm\pi/2$ near $k\approx 2\pi/d$.}
\label{fig:diff_cm}
\end{figure*}
The flux-normalized wave-function for mode $m$ in a lead of width $d$ oriented
parallel to the $x$-axis is
\begin{eqnarray}
\psi_m(x,y) & = & 
\sqrt{\frac{2}{dk_{x,m}}} 
e^{ik_{x,m}x} \sin\left(k_{y,m} y\right) \nonumber \\
& = & -i \sqrt{\frac{1}{2dk_{x,m}}}  [ 
e^{i(k_{x,m}x + k_{y,m}y)} \nonumber \\
&-&e^{i(k_{x,m}x- k_{y,m}y)}], 
\label{decomp}
\end{eqnarray}
where $k_{y,m} = m\pi/d$ is the transverse momentum component
and $k_{x,m} = \sqrt{k^2 - (m\pi/d)^2}$ the longitudinal component. Eq.~(\ref{decomp}) can be viewed as two rays emanating with angles
$\pm\theta_m =\pm\arcsin(m\pi/dk)$ (Fig.~\ref{fig:difflead}). 
In the GTD approximation, each ray hits an edge of the lead mouth 
and the two cylindrical 
waves emanating from the edges cause, in turn, an interference pattern at large distances.
The GTD diffraction for scattering from the lead mode $m$ into the
half-space (in our application the cavity) can be written in direct analogy with Eqs.~\ref{eq:vgtd}
and \ref{eq:gtdampl} as
\begin{eqnarray}
c_m^{\rm{GTD}}(\theta,k,d)&=&\frac{-ie^{\frac{im\pi}{2}}}{\sqrt{2d}k_{x,m}}\Big[\frac{1}{2}D_L(\theta,\theta_m)e^{i\frac{m\pi}{2}}e^{-ik\frac{d}2\sin\theta}\nonumber \\ &-&
\frac{1}{2}D_R(\theta,\theta_m)e^{-i\frac{m\pi}{2}}e^{ik\frac{d}2\sin\theta}\Big],
\label{eq:dngtd}
\end{eqnarray} 
with the scattering coefficients at left and right wedge
\begin{eqnarray}
D_L(\theta,\theta_m) & = & D(\frac{\pi}{2}-\theta,\frac{3\pi}{2}-\theta_m), 
\nonumber \\
D_R(\theta,\theta_m) & = & D(\frac{\pi}{2}+\theta,\frac{3\pi}{2}-\theta_m).
\end{eqnarray}
The two paths have
a phase difference of $m\pi$ at the lead mouth and thus a phase difference $\pm m\pi/2$
relative to the reference path that starts at the center 
of the orifice (see Fig.~\ref{fig:difflead}).
With the flux normalization 
factor $\sqrt{k_{x,m}}$ from Eq.~(\ref{eq:tnm}) $\sqrt{k_{x,m}}c_m^{\rm{GTD}}(\theta,k,d)$ is dimensionless .\\
We include now higher-order scattering events on the level of the 
UTD in order to improve Eq.~(\ref{eq:dngtd}), 
\begin{widetext}
\begin{eqnarray} 
c_m^{\rm{GTD-UTD}}(\theta,k,d)=c_m^{\rm{GTD}}(\theta,k,d) -i \frac{e^{\frac{im\pi}{2}}}{\sqrt{2d}k_{x,m}}&\times& \nonumber \\
\frac{1}{4}\Bigg[\sum_{\rm{odd}:j=1}^{j_{\rm max}}U_R(\theta,+\pi/2,jd,k)g_j(k)e^{i\phi_{++}}D_L(-\pi/2,\theta_m)-
U_L(\theta,-\pi/2,jd,k)g_j(k)e^{i\Phi_{--}}D_R(+\pi/2,\theta_m)&-& \nonumber \\
\sum_{\rm {even}:j=1}^{j_{\rm max}}U_R(\theta,+\pi/2,jd,k)g_j(k)e^{i\Phi_{+-}}D_R(+\pi/2,\theta_m)+
U_L(\theta,-\pi/2,jd,k)g_j(k)e^{i\Phi_{-+}}D_L(-\pi/2,\theta_m)\Bigg], \nonumber \\
\end{eqnarray}
\end{widetext}
where 
\begin{eqnarray}
\Phi_{\pm \pm}=\pm k\frac{d}{2}\sin{\theta}\pm \frac{m\pi}{2},
\end{eqnarray}
$g_j(k)$ is given in Eq.~(\ref{gkj}) and $U_L$, $U_R$ are given by Eq.~(\ref{eq:Utd}).
Fig.~\ref{fig:diff_cm} illustrates that the UTD corrections become most important
when the value of $k$ is close to a threshold,
i.e., when the angle $\theta_m$ is close to $\pi/2$.
%---------------------------------------------------------------------------------------------------
%Bibliography
%---------------------------------------------------------------------------------------------------

%------------------------------------------------------------------------------

\begin{thebibliography}{0}
\expandafter\ifx\csname natexlab\endcsname\relax\def\natexlab#1{#1}\fi
\expandafter\ifx\csname bibnamefont\endcsname\relax
  \def\bibnamefont#1{#1}\fi
\expandafter\ifx\csname bibfnamefont\endcsname\relax
  \def\bibfnamefont#1{#1}\fi
\expandafter\ifx\csname citenamefont\endcsname\relax
  \def\citenamefont#1{#1}\fi
\expandafter\ifx\csname url\endcsname\relax
  \def\url#1{\texttt{#1}}\fi
\expandafter\ifx\csname urlprefix\endcsname\relax\def\urlprefix{URL }\fi
\providecommand{\bibinfo}[2]{#2}
\providecommand{\eprint}[2][]{\url{#2}}

\end{thebibliography}


\begin{thebibliography}{63}
\expandafter\ifx\csname natexlab\endcsname\relax\def\natexlab#1{#1}\fi
\expandafter\ifx\csname bibnamefont\endcsname\relax
  \def\bibnamefont#1{#1}\fi
\expandafter\ifx\csname bibfnamefont\endcsname\relax
  \def\bibfnamefont#1{#1}\fi
\expandafter\ifx\csname citenamefont\endcsname\relax
  \def\citenamefont#1{#1}\fi
\expandafter\ifx\csname url\endcsname\relax
  \def\url#1{\texttt{#1}}\fi
\expandafter\ifx\csname urlprefix\endcsname\relax\def\urlprefix{URL }\fi
\providecommand{\bibinfo}[2]{#2}
\providecommand{\eprint}[2][]{\url{#2}}

\bibitem[{\citenamefont{Datta}(1995)}]{Dat95}
\bibinfo{author}{\bibfnamefont{S.}~\bibnamefont{Datta}},
  \emph{\bibinfo{title}{Electronic Transport in Mesoscopic Systems}}
  (\bibinfo{publisher}{Cambridge University Press}, \bibinfo{year}{1995}).

\bibitem[{\citenamefont{Akkermans and Montambaux}(2006)}]{AkkMon06}
\bibinfo{author}{\bibfnamefont{E.}~\bibnamefont{Akkermans}} \bibnamefont{and}
  \bibinfo{author}{\bibfnamefont{G.}~\bibnamefont{Montambaux}},
  \emph{\bibinfo{title}{Mesoscopic Physics of Electrons and Photons}}
  (\bibinfo{publisher}{Cambridge University Press}, \bibinfo{year}{2006}).

\bibitem[{\citenamefont{Beenakker and van Houten}(1991)}]{bee91}
\bibinfo{author}{\bibfnamefont{C.~W.~J.} \bibnamefont{Beenakker}}
  \bibnamefont{and} \bibinfo{author}{\bibfnamefont{H.}~\bibnamefont{van
  Houten}}, \bibinfo{journal}{Solid State Physics}
  \textbf{\bibinfo{volume}{4}}, \bibinfo{pages}{1} (\bibinfo{year}{1991}).

\bibitem[{\citenamefont{Gutzwiller}(1991)}]{gutz91}
\bibinfo{author}{\bibfnamefont{M.~C.} \bibnamefont{Gutzwiller}},
  \emph{\bibinfo{title}{Chaos in Classical and Quantum Mechanics}}
  (\bibinfo{publisher}{Springer Verlag, New York}, \bibinfo{year}{1991}).

\bibitem[{\citenamefont{Berry and Mount}(1972)}]{ber72}
\bibinfo{author}{\bibfnamefont{N.~V.} \bibnamefont{Berry}} \bibnamefont{and}
  \bibinfo{author}{\bibfnamefont{K.~E.} \bibnamefont{Mount}},
  \bibinfo{journal}{Rep. Prog. Phys.} \textbf{\bibinfo{volume}{35}},
  \bibinfo{pages}{315} (\bibinfo{year}{1972}).

\bibitem[{\citenamefont{Brack and Bhaduri}(2003)}]{BraBha03}
\bibinfo{author}{\bibfnamefont{M.}~\bibnamefont{Brack}} \bibnamefont{and}
  \bibinfo{author}{\bibfnamefont{R.~K.} \bibnamefont{Bhaduri}},
  \emph{\bibinfo{title}{Semiclassical Physics}} (\bibinfo{publisher}{Frontiers
  in Physics, Westview Press}, \bibinfo{year}{2003}).

\bibitem[{\citenamefont{Bl\"umel and Smilansky}(1990)}]{BluSmi90}
\bibinfo{author}{\bibfnamefont{R.}~\bibnamefont{Bl\"umel}} \bibnamefont{and}
  \bibinfo{author}{\bibfnamefont{U.}~\bibnamefont{Smilansky}},
  \bibinfo{journal}{Phys. Rev. Lett.} \textbf{\bibinfo{volume}{64}},
  \bibinfo{pages}{241} (\bibinfo{year}{1990}).

\bibitem[{\citenamefont{Baranger et~al.}(1993)\citenamefont{Baranger, Jalabert,
  and Stone}}]{BarJalSto93}
\bibinfo{author}{\bibfnamefont{H.~U.} \bibnamefont{Baranger}},
  \bibinfo{author}{\bibfnamefont{R.~A.} \bibnamefont{Jalabert}},
  \bibnamefont{and} \bibinfo{author}{\bibfnamefont{A.~D.} \bibnamefont{Stone}},
  \bibinfo{journal}{Chaos} \textbf{\bibinfo{volume}{3}}, \bibinfo{pages}{665}
  (\bibinfo{year}{1993}).

\bibitem[{\citenamefont{Blomquist}(2002)}]{blom02}
\bibinfo{author}{\bibfnamefont{T.}~\bibnamefont{Blomquist}},
  \bibinfo{journal}{Phys. Rev. B} \textbf{\bibinfo{volume}{66}},
  \bibinfo{pages}{155316} (\bibinfo{year}{2002}).

\bibitem[{\citenamefont{Schwieters et~al.}(1996)\citenamefont{Schwieters,
  Alford, and Delos}}]{SchAlfDel96}
\bibinfo{author}{\bibfnamefont{C.~D.} \bibnamefont{Schwieters}},
  \bibinfo{author}{\bibfnamefont{J.~A.} \bibnamefont{Alford}},
  \bibnamefont{and} \bibinfo{author}{\bibfnamefont{J.~B.} \bibnamefont{Delos}},
  \bibinfo{journal}{Phys. Rev. B} \textbf{\bibinfo{volume}{54}},
  \bibinfo{pages}{10652} (\bibinfo{year}{1996}).

\bibitem[{\citenamefont{Blomquist and Zozoulenko}(2001)}]{BloZoz01}
\bibinfo{author}{\bibfnamefont{T.}~\bibnamefont{Blomquist}} \bibnamefont{and}
  \bibinfo{author}{\bibfnamefont{I.~V.} \bibnamefont{Zozoulenko}},
  \bibinfo{journal}{Phys. Rev. B} \textbf{\bibinfo{volume}{64}},
  \bibinfo{pages}{195301} (\bibinfo{year}{2001}).

\bibitem[{\citenamefont{Wirtz et~al.}(1997)\citenamefont{Wirtz, Tang, and
  Burgd\"orfer}}]{WirTanBur97}
\bibinfo{author}{\bibfnamefont{L.}~\bibnamefont{Wirtz}},
  \bibinfo{author}{\bibfnamefont{J.-Z.} \bibnamefont{Tang}}, \bibnamefont{and}
  \bibinfo{author}{\bibfnamefont{J.}~\bibnamefont{Burgd\"orfer}},
  \bibinfo{journal}{Phys. Rev. B} \textbf{\bibinfo{volume}{56}},
  \bibinfo{pages}{7589} (\bibinfo{year}{1997}).

\bibitem[{\citenamefont{Wirtz et~al.}(1999)\citenamefont{Wirtz, Tang, and
  Burgd\"orfer}}]{WirTanBur99}
\bibinfo{author}{\bibfnamefont{L.}~\bibnamefont{Wirtz}},
  \bibinfo{author}{\bibfnamefont{J.-Z.} \bibnamefont{Tang}}, \bibnamefont{and}
  \bibinfo{author}{\bibfnamefont{J.}~\bibnamefont{Burgd\"orfer}},
  \bibinfo{journal}{Phys. Rev. B} \textbf{\bibinfo{volume}{59}},
  \bibinfo{pages}{2956} (\bibinfo{year}{1999}).

\bibitem[{\citenamefont{Wirtz et~al.}(2003)\citenamefont{Wirtz, Stampfer,
  Rotter, and Burgd\"orfer}}]{WirStaRotBur03}
\bibinfo{author}{\bibfnamefont{L.}~\bibnamefont{Wirtz}},
  \bibinfo{author}{\bibfnamefont{C.}~\bibnamefont{Stampfer}},
  \bibinfo{author}{\bibfnamefont{S.}~\bibnamefont{Rotter}}, \bibnamefont{and}
  \bibinfo{author}{\bibfnamefont{J.}~\bibnamefont{Burgd\"orfer}},
  \bibinfo{journal}{Phys. Rev. E} \textbf{\bibinfo{volume}{67}},
  \bibinfo{pages}{016206} (\bibinfo{year}{2003}).

\bibitem[{\citenamefont{Stampfer et~al.}(2005)\citenamefont{Stampfer, Rotter,
  Burgd\"orfer, and Wirtz}}]{StaRotBurWir05}
\bibinfo{author}{\bibfnamefont{C.}~\bibnamefont{Stampfer}},
  \bibinfo{author}{\bibfnamefont{S.}~\bibnamefont{Rotter}},
  \bibinfo{author}{\bibfnamefont{J.}~\bibnamefont{Burgd\"orfer}},
  \bibnamefont{and} \bibinfo{author}{\bibfnamefont{L.}~\bibnamefont{Wirtz}},
  \bibinfo{journal}{Phys. Rev. E} \textbf{\bibinfo{volume}{72}},
  \bibinfo{pages}{036223} (\bibinfo{year}{2005}).

\bibitem[{\citenamefont{B\v{r}ezinov\'{a}
  et~al.}(2008)\citenamefont{B\v{r}ezinov\'{a}, Stampfer, Wirtz, Rotter, and
  Burgd\"{o}rfer}}]{BreStaWirRotBur08}
\bibinfo{author}{\bibfnamefont{I.}~\bibnamefont{B\v{r}ezinov\'{a}}},
  \bibinfo{author}{\bibfnamefont{C.}~\bibnamefont{Stampfer}},
  \bibinfo{author}{\bibfnamefont{L.}~\bibnamefont{Wirtz}},
  \bibinfo{author}{\bibfnamefont{S.}~\bibnamefont{Rotter}}, \bibnamefont{and}
  \bibinfo{author}{\bibfnamefont{J.}~\bibnamefont{Burgd\"{o}rfer}},
  \bibinfo{journal}{Phys. Rev. B} \textbf{\bibinfo{volume}{77}},
  \bibinfo{pages}{165321} (\bibinfo{year}{2008}).

\bibitem[{\citenamefont{Rahav and Brouwer}(2005)}]{RahBro05}
\bibinfo{author}{\bibfnamefont{S.}~\bibnamefont{Rahav}} \bibnamefont{and}
  \bibinfo{author}{\bibfnamefont{P.~W.} \bibnamefont{Brouwer}},
  \bibinfo{journal}{Phys. Rev. Lett.} \textbf{\bibinfo{volume}{95}},
  \bibinfo{pages}{056806} (\bibinfo{year}{2005}).

\bibitem[{\citenamefont{Rahav and Brouwer}(2006)}]{RahBro06}
\bibinfo{author}{\bibfnamefont{S.}~\bibnamefont{Rahav}} \bibnamefont{and}
  \bibinfo{author}{\bibfnamefont{P.~W.} \bibnamefont{Brouwer}},
  \bibinfo{journal}{Phys. Rev. B} \textbf{\bibinfo{volume}{73}},
  \bibinfo{eid}{035324} (\bibinfo{year}{2006}).

\bibitem[{\citenamefont{Jacquod and Whitney}(2006)}]{JacWhi06}
\bibinfo{author}{\bibfnamefont{P.}~\bibnamefont{Jacquod}} \bibnamefont{and}
  \bibinfo{author}{\bibfnamefont{R.~S.} \bibnamefont{Whitney}},
  \bibinfo{journal}{Phys. Rev. B} \textbf{\bibinfo{volume}{73}},
  \bibinfo{pages}{195115} (\bibinfo{year}{2006}).

\bibitem[{\citenamefont{Richter and Sieber}(2002)}]{RicSie02}
\bibinfo{author}{\bibfnamefont{K.}~\bibnamefont{Richter}} \bibnamefont{and}
  \bibinfo{author}{\bibfnamefont{M.}~\bibnamefont{Sieber}},
  \bibinfo{journal}{Phys. Rev. Lett.} \textbf{\bibinfo{volume}{89}},
  \bibinfo{pages}{206801} (\bibinfo{year}{2002}).

\bibitem[{\citenamefont{Braun et~al.}(2006)\citenamefont{Braun, Heusler,
  M\"uller, and Haake}}]{BraHeuMulHaa06}
\bibinfo{author}{\bibfnamefont{P.}~\bibnamefont{Braun}},
  \bibinfo{author}{\bibfnamefont{S.}~\bibnamefont{Heusler}},
  \bibinfo{author}{\bibfnamefont{S.}~\bibnamefont{M\"uller}}, \bibnamefont{and}
  \bibinfo{author}{\bibfnamefont{F.}~\bibnamefont{Haake}}, \bibinfo{journal}{J.
  Phys. A.: Math. Gen.} \textbf{\bibinfo{volume}{39}}, \bibinfo{pages}{L159}
  (\bibinfo{year}{2006}).

\bibitem[{\citenamefont{Heusler et~al.}(2006)\citenamefont{Heusler, M\"{u}ller,
  Braun, and Haake}}]{HeuMulBraHaa06}
\bibinfo{author}{\bibfnamefont{S.}~\bibnamefont{Heusler}},
  \bibinfo{author}{\bibfnamefont{S.}~\bibnamefont{M\"{u}ller}},
  \bibinfo{author}{\bibfnamefont{P.}~\bibnamefont{Braun}}, \bibnamefont{and}
  \bibinfo{author}{\bibfnamefont{F.}~\bibnamefont{Haake}},
  \bibinfo{journal}{Phys. Rev. Lett.} \textbf{\bibinfo{volume}{96}},
  \bibinfo{pages}{066804} (\bibinfo{year}{2006}).

\bibitem[{\citenamefont{Brouwer and Rahav}(2006)}]{BroRah06}
\bibinfo{author}{\bibfnamefont{P.~W.} \bibnamefont{Brouwer}} \bibnamefont{and}
  \bibinfo{author}{\bibfnamefont{S.}~\bibnamefont{Rahav}},
  \bibinfo{journal}{Phys. Rev. B} \textbf{\bibinfo{volume}{74}},
  \bibinfo{pages}{085313} (\bibinfo{year}{2006}).

\bibitem[{\citenamefont{Pichaureau and Jalabert}(1999)}]{PicJal99}
\bibinfo{author}{\bibfnamefont{P.}~\bibnamefont{Pichaureau}} \bibnamefont{and}
  \bibinfo{author}{\bibfnamefont{R.~A.} \bibnamefont{Jalabert}},
  \bibinfo{journal}{European Physical Journal B} \textbf{\bibinfo{volume}{9}},
  \bibinfo{pages}{299} (\bibinfo{year}{1999}).

\bibitem[{\citenamefont{Bogomolny}(2000)}]{bog00}
\bibinfo{author}{\bibfnamefont{E.}~\bibnamefont{Bogomolny}},
  \bibinfo{journal}{Nonlinearity} \textbf{\bibinfo{volume}{13}},
  \bibinfo{pages}{947} (\bibinfo{year}{2000}).

\bibitem[{\citenamefont{Argaman}(1996)}]{arg96b}
\bibinfo{author}{\bibfnamefont{N.}~\bibnamefont{Argaman}},
  \bibinfo{journal}{Phys. Rev. B} \textbf{\bibinfo{volume}{53}},
  \bibinfo{pages}{7035} (\bibinfo{year}{1996}).

\bibitem[{\citenamefont{Feynman and Hibbs}(1965)}]{fey65}
\bibinfo{author}{\bibfnamefont{R.~P.} \bibnamefont{Feynman}} \bibnamefont{and}
  \bibinfo{author}{\bibfnamefont{A.~R.} \bibnamefont{Hibbs}},
  \emph{\bibinfo{title}{Quantum Mechanics and Path Integrals}}
  (\bibinfo{publisher}{MacGraw-Hill, New York}, \bibinfo{year}{1965}).

\bibitem[{\citenamefont{Chang et~al.}(1994)\citenamefont{Chang, Baranger,
  Pfeiffer, and West}}]{ChaBarPfeWes94}
\bibinfo{author}{\bibfnamefont{A.~M.} \bibnamefont{Chang}},
  \bibinfo{author}{\bibfnamefont{H.~U.} \bibnamefont{Baranger}},
  \bibinfo{author}{\bibfnamefont{L.~N.} \bibnamefont{Pfeiffer}},
  \bibnamefont{and} \bibinfo{author}{\bibfnamefont{K.~W.} \bibnamefont{West}},
  \bibinfo{journal}{Phys. Rev. Lett.} \textbf{\bibinfo{volume}{73}},
  \bibinfo{pages}{2111} (\bibinfo{year}{1994}).

\bibitem[{\citenamefont{Marcus et~al.}(1992)\citenamefont{Marcus, Rimberg,
  Westervelt, Hopkins, and Gossard}}]{MarRimWesHopGos92}
\bibinfo{author}{\bibfnamefont{C.~M.} \bibnamefont{Marcus}},
  \bibinfo{author}{\bibfnamefont{A.~J.} \bibnamefont{Rimberg}},
  \bibinfo{author}{\bibfnamefont{R.~M.} \bibnamefont{Westervelt}},
  \bibinfo{author}{\bibfnamefont{P.~F.} \bibnamefont{Hopkins}},
  \bibnamefont{and} \bibinfo{author}{\bibfnamefont{A.~C.}
  \bibnamefont{Gossard}}, \bibinfo{journal}{Phys. Rev. Lett.}
  \textbf{\bibinfo{volume}{69}}, \bibinfo{pages}{506} (\bibinfo{year}{1992}).

\bibitem[{\citenamefont{Kouyoumjian and Pathak}(1974)}]{KouPat74}
\bibinfo{author}{\bibfnamefont{R.}~\bibnamefont{Kouyoumjian}} \bibnamefont{and}
  \bibinfo{author}{\bibfnamefont{P.~H.} \bibnamefont{Pathak}},
  \bibinfo{journal}{Proc. IEEE} \textbf{\bibinfo{volume}{62}},
  \bibinfo{pages}{1448} (\bibinfo{year}{1974}).

\bibitem[{\citenamefont{Sieber et~al.}(1997)\citenamefont{Sieber, Pavloff, and
  Schmit}}]{SiePavSch97}
\bibinfo{author}{\bibfnamefont{M.}~\bibnamefont{Sieber}},
  \bibinfo{author}{\bibfnamefont{N.}~\bibnamefont{Pavloff}}, \bibnamefont{and}
  \bibinfo{author}{\bibfnamefont{C.}~\bibnamefont{Schmit}},
  \bibinfo{journal}{Phys. Rev. E} \textbf{\bibinfo{volume}{55}},
  \bibinfo{pages}{2279} (\bibinfo{year}{1997}).

\bibitem[{\citenamefont{Keller}(1962)}]{Kel62}
\bibinfo{author}{\bibfnamefont{J.~B.} \bibnamefont{Keller}},
  \bibinfo{journal}{J. Opt. Soc. Am} \textbf{\bibinfo{volume}{52}},
  \bibinfo{pages}{116} (\bibinfo{year}{1962}).

\bibitem[{\citenamefont{Rotter et~al.}(2000)\citenamefont{Rotter, Tang, Wirtz,
  Trost, and Burgd\"orfer}}]{RotTanWir00}
\bibinfo{author}{\bibfnamefont{S.}~\bibnamefont{Rotter}},
  \bibinfo{author}{\bibfnamefont{J.-Z.} \bibnamefont{Tang}},
  \bibinfo{author}{\bibfnamefont{L.}~\bibnamefont{Wirtz}},
  \bibinfo{author}{\bibfnamefont{J.}~\bibnamefont{Trost}}, \bibnamefont{and}
  \bibinfo{author}{\bibfnamefont{J.}~\bibnamefont{Burgd\"orfer}},
  \bibinfo{journal}{Phys. Rev. B} \textbf{\bibinfo{volume}{62}},
  \bibinfo{pages}{1950} (\bibinfo{year}{2000}).

\bibitem[{\citenamefont{Rotter et~al.}(2003)\citenamefont{Rotter, Weingartner,
  Rohringer, and Burgd\"orfer}}]{RotWeiRohBur03}
\bibinfo{author}{\bibfnamefont{S.}~\bibnamefont{Rotter}},
  \bibinfo{author}{\bibfnamefont{B.}~\bibnamefont{Weingartner}},
  \bibinfo{author}{\bibfnamefont{N.}~\bibnamefont{Rohringer}},
  \bibnamefont{and}
  \bibinfo{author}{\bibfnamefont{J.}~\bibnamefont{Burgd\"orfer}},
  \bibinfo{journal}{Phys. Rev. B} \textbf{\bibinfo{volume}{68}},
  \bibinfo{pages}{165302} (\bibinfo{year}{2003}).

\bibitem[{\citenamefont{Ishio and Burgd\"orfer}(1995)}]{IshBur95}
\bibinfo{author}{\bibfnamefont{H.}~\bibnamefont{Ishio}} \bibnamefont{and}
  \bibinfo{author}{\bibfnamefont{J.}~\bibnamefont{Burgd\"orfer}},
  \bibinfo{journal}{Phys. Rev. B} \textbf{\bibinfo{volume}{51}},
  \bibinfo{pages}{2013} (\bibinfo{year}{1995}).

\bibitem[{\citenamefont{Fisher and Lee}(1981)}]{FisLee81}
\bibinfo{author}{\bibfnamefont{D.~S.} \bibnamefont{Fisher}} \bibnamefont{and}
  \bibinfo{author}{\bibfnamefont{P.~A.} \bibnamefont{Lee}},
  \bibinfo{journal}{Phys. Rev. B} \textbf{\bibinfo{volume}{23}},
  \bibinfo{pages}{6851} (\bibinfo{year}{1981}).

\bibitem[{\citenamefont{Szeredi and Goodings}(1993)}]{SzeGoo93}
\bibinfo{author}{\bibfnamefont{T.}~\bibnamefont{Szeredi}} \bibnamefont{and}
  \bibinfo{author}{\bibfnamefont{D.~A.} \bibnamefont{Goodings}},
  \bibinfo{journal}{Phys. Rev. E} \textbf{\bibinfo{volume}{48}},
  \bibinfo{pages}{3529} (\bibinfo{year}{1993}).

\bibitem[{\citenamefont{Hikami}(1981)}]{Hik81}
\bibinfo{author}{\bibfnamefont{S.}~\bibnamefont{Hikami}},
  \bibinfo{journal}{Phys. Rev. B} \textbf{\bibinfo{volume}{24}},
  \bibinfo{pages}{2671} (\bibinfo{year}{1981}).

\bibitem[{\citenamefont{Andreev}(1964)}]{And64}
\bibinfo{author}{\bibfnamefont{A.}~\bibnamefont{Andreev}},
  \bibinfo{journal}{Sov. Phys. JETP} \textbf{\bibinfo{volume}{19}},
  \bibinfo{pages}{1228} (\bibinfo{year}{1964}).

\bibitem[{\citenamefont{Yang et~al.}(1995)\citenamefont{Yang, Ishio, and
  Burgd\"{o}rfer}}]{yan95}
\bibinfo{author}{\bibfnamefont{X.}~\bibnamefont{Yang}},
  \bibinfo{author}{\bibfnamefont{H.}~\bibnamefont{Ishio}}, \bibnamefont{and}
  \bibinfo{author}{\bibfnamefont{J.}~\bibnamefont{Burgd\"{o}rfer}},
  \bibinfo{journal}{Phys. Rev. B} \textbf{\bibinfo{volume}{52}},
  \bibinfo{pages}{8219} (\bibinfo{year}{1995}).

\bibitem[{\citenamefont{Tworzyd\l{}o et~al.}(2004)\citenamefont{Tworzyd\l{}o,
  Tajic, and Beenakker}}]{TwoTajBee04}
\bibinfo{author}{\bibfnamefont{J.}~\bibnamefont{Tworzyd\l{}o}},
  \bibinfo{author}{\bibfnamefont{A.}~\bibnamefont{Tajic}}, \bibnamefont{and}
  \bibinfo{author}{\bibfnamefont{C.~W.~J.} \bibnamefont{Beenakker}},
  \bibinfo{journal}{Phys. Rev. B} \textbf{\bibinfo{volume}{69}},
  \bibinfo{pages}{165318} (\bibinfo{year}{2004}).

\bibitem[{\citenamefont{Jacquod and Sukhorukov}(2004)}]{JacSuk04}
\bibinfo{author}{\bibfnamefont{P.}~\bibnamefont{Jacquod}} \bibnamefont{and}
  \bibinfo{author}{\bibfnamefont{E.~V.} \bibnamefont{Sukhorukov}},
  \bibinfo{journal}{Phys. Rev. Lett.} \textbf{\bibinfo{volume}{92}},
  \bibinfo{pages}{116801} (\bibinfo{year}{2004}).

\bibitem[{\citenamefont{Kharitonov and Efetov}(2008)}]{KhaEfe08}
\bibinfo{author}{\bibfnamefont{M.~Y.} \bibnamefont{Kharitonov}}
  \bibnamefont{and} \bibinfo{author}{\bibfnamefont{K.~B.}
  \bibnamefont{Efetov}}, \bibinfo{journal}{Phys. Rev. B}
  \textbf{\bibinfo{volume}{78}}, \bibinfo{pages}{033404}
  (\bibinfo{year}{2008}).

\bibitem[{\citenamefont{Weingartner et~al.}(2005)\citenamefont{Weingartner,
  Rotter, and Burgd\"orfer}}]{WeiRotBur05}
\bibinfo{author}{\bibfnamefont{B.}~\bibnamefont{Weingartner}},
  \bibinfo{author}{\bibfnamefont{S.}~\bibnamefont{Rotter}}, \bibnamefont{and}
  \bibinfo{author}{\bibfnamefont{J.}~\bibnamefont{Burgd\"orfer}},
  \bibinfo{journal}{Phys. Rev. B} \textbf{\bibinfo{volume}{72}},
  \bibinfo{pages}{115342} (\bibinfo{year}{2005}).

\bibitem[{\citenamefont{Hartung et~al.}(2008)\citenamefont{Hartung, Wellens,
  M\"{u}ller, Richter, and Schlagheck}}]{HarWelMulRicSch08}
\bibinfo{author}{\bibfnamefont{M.}~\bibnamefont{Hartung}},
  \bibinfo{author}{\bibfnamefont{T.}~\bibnamefont{Wellens}},
  \bibinfo{author}{\bibfnamefont{C.~A.} \bibnamefont{M\"{u}ller}},
  \bibinfo{author}{\bibfnamefont{K.}~\bibnamefont{Richter}}, \bibnamefont{and}
  \bibinfo{author}{\bibfnamefont{P.}~\bibnamefont{Schlagheck}},
  \bibinfo{journal}{Phys. Rev. Lett.} \textbf{\bibinfo{volume}{101}},
  \bibinfo{pages}{020603} (\bibinfo{year}{2008}).

\bibitem[{\citenamefont{Larose et~al.}(2004)\citenamefont{Larose, Margerin, van
  Tiggelen, and Campillo}}]{LarMarTigCam04}
\bibinfo{author}{\bibfnamefont{E.}~\bibnamefont{Larose}},
  \bibinfo{author}{\bibfnamefont{L.}~\bibnamefont{Margerin}},
  \bibinfo{author}{\bibfnamefont{B.~A.} \bibnamefont{van Tiggelen}},
  \bibnamefont{and} \bibinfo{author}{\bibfnamefont{M.}~\bibnamefont{Campillo}},
  \bibinfo{journal}{Phys. Rev. Lett.} \textbf{\bibinfo{volume}{93}},
  \bibinfo{pages}{048501} (\bibinfo{year}{2004}).

\bibitem[{\citenamefont{Robinson et~al.}(2008)\citenamefont{Robinson,
  Schomerus, Oroszl\'{a}ny, and Fal'ko}}]{RobSchOroFal08}
\bibinfo{author}{\bibfnamefont{J.~P.} \bibnamefont{Robinson}},
  \bibinfo{author}{\bibfnamefont{H.}~\bibnamefont{Schomerus}},
  \bibinfo{author}{\bibfnamefont{L.}~\bibnamefont{Oroszl\'{a}ny}},
  \bibnamefont{and} \bibinfo{author}{\bibfnamefont{V.~I.}
  \bibnamefont{Fal'ko}}, \bibinfo{journal}{Phys. Rev. Lett.}
  \textbf{\bibinfo{volume}{101}}, \bibinfo{pages}{196803}
  (\bibinfo{year}{2008}).

\bibitem[{\citenamefont{Kopp et~al.}(2008)\citenamefont{Kopp, Schomerus, and
  Rotter}}]{KopSchRot08}
\bibinfo{author}{\bibfnamefont{M.}~\bibnamefont{Kopp}},
  \bibinfo{author}{\bibfnamefont{H.}~\bibnamefont{Schomerus}},
  \bibnamefont{and} \bibinfo{author}{\bibfnamefont{S.}~\bibnamefont{Rotter}},
  \bibinfo{journal}{Phys. Rev. B} \textbf{\bibinfo{volume}{78}},
  \bibinfo{pages}{075312} (\bibinfo{year}{2008}).

\bibitem[{\citenamefont{Gopar et~al.}(2006)\citenamefont{Gopar, Rotter, and
  Schomerus}}]{GopRotSch06}
\bibinfo{author}{\bibfnamefont{V.~A.} \bibnamefont{Gopar}},
  \bibinfo{author}{\bibfnamefont{S.}~\bibnamefont{Rotter}}, \bibnamefont{and}
  \bibinfo{author}{\bibfnamefont{H.}~\bibnamefont{Schomerus}},
  \bibinfo{journal}{Phys. Rev. B} \textbf{\bibinfo{volume}{73}},
  \bibinfo{pages}{165308} (\bibinfo{year}{2006}).

\bibitem[{\citenamefont{Blanter and B\"{u}ttiker}(2000)}]{BlaBut00}
\bibinfo{author}{\bibfnamefont{Y.~M.} \bibnamefont{Blanter}} \bibnamefont{and}
  \bibinfo{author}{\bibfnamefont{M.}~\bibnamefont{B\"{u}ttiker}},
  \bibinfo{journal}{Phys. Rep.} \textbf{\bibinfo{volume}{336}},
  \bibinfo{pages}{1} (\bibinfo{year}{2000}).

\bibitem[{\citenamefont{Aigner et~al.}(2005)\citenamefont{Aigner, Rotter, and
  Burgd\"orfer}}]{AigRotBur05}
\bibinfo{author}{\bibfnamefont{F.}~\bibnamefont{Aigner}},
  \bibinfo{author}{\bibfnamefont{S.}~\bibnamefont{Rotter}}, \bibnamefont{and}
  \bibinfo{author}{\bibfnamefont{J.}~\bibnamefont{Burgd\"orfer}},
  \bibinfo{journal}{Phys. Rev. Lett.} \textbf{\bibinfo{volume}{94}},
  \bibinfo{pages}{216801} (\bibinfo{year}{2005}).

\bibitem[{\citenamefont{Rotter et~al.}(2007)\citenamefont{Rotter, Aigner, and
  Burgd\"{o}rfer}}]{RotAigBur07}
\bibinfo{author}{\bibfnamefont{S.}~\bibnamefont{Rotter}},
  \bibinfo{author}{\bibfnamefont{F.}~\bibnamefont{Aigner}}, \bibnamefont{and}
  \bibinfo{author}{\bibfnamefont{J.}~\bibnamefont{Burgd\"{o}rfer}},
  \bibinfo{journal}{Phys. Rev. B} \textbf{\bibinfo{volume}{75}},
  \bibinfo{pages}{125312} (\bibinfo{year}{2007}).

\bibitem[{\citenamefont{Khoruzhenko et~al.}(2009)\citenamefont{Khoruzhenko,
  Savin, and Sommers}}]{KhoSavSom09}
\bibinfo{author}{\bibfnamefont{B.~A.} \bibnamefont{Khoruzhenko}},
  \bibinfo{author}{\bibfnamefont{D.~V.} \bibnamefont{Savin}}, \bibnamefont{and}
  \bibinfo{author}{\bibfnamefont{H.-J.} \bibnamefont{Sommers}},
  \bibinfo{journal}{Phys. Rev. B} \textbf{\bibinfo{volume}{80}},
  \bibinfo{pages}{125301} (\bibinfo{year}{2009}).

\bibitem[{\citenamefont{Novaes}(2007)}]{Nov07}
\bibinfo{author}{\bibfnamefont{M.}~\bibnamefont{Novaes}},
  \bibinfo{journal}{Phys. Rev. B} \textbf{\bibinfo{volume}{75}},
  \bibinfo{pages}{073304} (\bibinfo{year}{2007}).

\bibitem[{\citenamefont{Tworzyd\l{}o et~al.}(2003)\citenamefont{Tworzyd\l{}o,
  Tajic, Schomerus, and Beenakker}}]{TwoTajSchBee03}
\bibinfo{author}{\bibfnamefont{J.}~\bibnamefont{Tworzyd\l{}o}},
  \bibinfo{author}{\bibfnamefont{A.}~\bibnamefont{Tajic}},
  \bibinfo{author}{\bibfnamefont{H.}~\bibnamefont{Schomerus}},
  \bibnamefont{and} \bibinfo{author}{\bibfnamefont{C.~W.~J.}
  \bibnamefont{Beenakker}}, \bibinfo{journal}{Phys. Rev. B}
  \textbf{\bibinfo{volume}{68}}, \bibinfo{pages}{115313}
  (\bibinfo{year}{2003}).

\bibitem[{\citenamefont{Lewenkopf et~al.}(2008)\citenamefont{Lewenkopf,
  Mucciolo, and CastroNeto}}]{LewMucCas08}
\bibinfo{author}{\bibfnamefont{C.~H.} \bibnamefont{Lewenkopf}},
  \bibinfo{author}{\bibfnamefont{E.~R.} \bibnamefont{Mucciolo}},
  \bibnamefont{and} \bibinfo{author}{\bibfnamefont{A.~H.}
  \bibnamefont{CastroNeto}}, \bibinfo{journal}{Phys. Rev. B}
  \textbf{\bibinfo{volume}{77}}, \bibinfo{pages}{081410(R)}
  (\bibinfo{year}{2008}).

\bibitem[{\citenamefont{Tworzyd\l{}o et~al.}(2006)\citenamefont{Tworzyd\l{}o,
  Trauzettel, Titov, Rycerz, and Beenakker}}]{TwoTraTitRycBee06}
\bibinfo{author}{\bibfnamefont{J.}~\bibnamefont{Tworzyd\l{}o}},
  \bibinfo{author}{\bibfnamefont{B.}~\bibnamefont{Trauzettel}},
  \bibinfo{author}{\bibfnamefont{M.}~\bibnamefont{Titov}},
  \bibinfo{author}{\bibfnamefont{A.}~\bibnamefont{Rycerz}}, \bibnamefont{and}
  \bibinfo{author}{\bibfnamefont{C.~W.~J.} \bibnamefont{Beenakker}},
  \bibinfo{journal}{Phys. Rev. Lett.} \textbf{\bibinfo{volume}{96}},
  \bibinfo{pages}{246802} (\bibinfo{year}{2006}).

\bibitem[{\citenamefont{Sommerfeld}(1950)}]{Som50}
\bibinfo{author}{\bibfnamefont{A.}~\bibnamefont{Sommerfeld}},
  \emph{\bibinfo{title}{Vorlesungen \"uber Theoretische Physik, Optik}}
  (\bibinfo{publisher}{Dieterich'sche Verlagsbuchhandlung},
  \bibinfo{year}{1950}).

\bibitem[{\citenamefont{Schneider and Luebbers}(1991)}]{SchLue91}
\bibinfo{author}{\bibfnamefont{M.}~\bibnamefont{Schneider}} \bibnamefont{and}
  \bibinfo{author}{\bibfnamefont{R.~J.} \bibnamefont{Luebbers}},
  \bibinfo{journal}{IEEE Trans. Antennas Propag.}
  \textbf{\bibinfo{volume}{39}}, \bibinfo{pages}{8} (\bibinfo{year}{1991}).

\bibitem[{\citenamefont{Schneider}(1988)}]{Sch88}
\bibinfo{author}{\bibfnamefont{M.}~\bibnamefont{Schneider}},
  \emph{\bibinfo{title}{A uniform solution of double wedge-edge diffraction}}
  (\bibinfo{publisher}{Ph.D.~thesis, Pennsylvania State University},
  \bibinfo{year}{1988}).

\bibitem[{\citenamefont{Albani}(2005)}]{Alb05}
\bibinfo{author}{\bibfnamefont{M.}~\bibnamefont{Albani}},
  \bibinfo{journal}{IEEE Trans. Antennas Propag.}
  \textbf{\bibinfo{volume}{53}}, \bibinfo{pages}{702} (\bibinfo{year}{2005}).

\bibitem[{\citenamefont{Holm}(1996)}]{Hol96}
\bibinfo{author}{\bibfnamefont{P.~D.} \bibnamefont{Holm}},
  \bibinfo{journal}{IEEE Trans. Antennas Propag.}
  \textbf{\bibinfo{volume}{44}}, \bibinfo{pages}{879} (\bibinfo{year}{1996}).

\bibitem[{\citenamefont{Reiche}(1911)}]{Rei11}
\bibinfo{author}{\bibfnamefont{F.}~\bibnamefont{Reiche}},
  \bibinfo{journal}{Annalen der Physik} \textbf{\bibinfo{volume}{37}},
  \bibinfo{pages}{131} (\bibinfo{year}{1911}).

\end{thebibliography}
\end{document}